%% file: 00_cercos.tex
\pdfoutput=1
\documentclass[review]{elsarticle}

\usepackage{amsmath}

\journal{European Journal of Mechanics - B/Fluids}

\usepackage{amssymb}
\usepackage{graphicx}% Include figure files
\usepackage{dcolumn}% Align table columns on decimal point
\usepackage[final]{changes}

\usepackage{subcaption}

\newlength{\figwidth}
\newlength{\allwidth}
\newcommand{\dsty}[1]{\displaystyle{#1}}
\newcommand{\bs}[1]{\boldsymbol{#1}}

\newcommand{\D}{\mathrm{d}}

\newcommand{\gradient}{\nabla}

\newcommand{\Grad}{\nabla}
\newcommand{\Div}[1]{\nabla \cdot {#1}}
\newcommand{\Lap}{\Delta}
\newcommand{\SPH}[1]{\left\langle {#1} \right\rangle}
\newcommand{\kin}{_{\mathrm{k}}}

\newcommand{\sph}[1]{\left\langle {#1} \right\rangle}
\newcommand{\vrate}[3]{\left( \dfrac{\Delta {#1}}{\Delta t}\right)_{#2}^{{#3}^{\ast}}}
\newcommand{\vrateni}[2]{\left( \dfrac{\Delta {#1}}{\Delta t}\right)^{{#2}^{\ast}}}

\begin{document}
	
\begin{frontmatter}
	
%\preprint{}

\input{01_title}

\date{\today}
\input{02_abstract}

\end{frontmatter}
% \linenumbers

%\maketitle

%\newpage
%\TOC

\input{10_intro}

\input{20_spatial_discretization}
\input{21_sph}
\input{22_sph_energy_balance}

\input{30_time_discretization}
\input{31_time_scheme}
\input{32_total_energy_balance}

\input{33_implicit_time_step}

\input{40_practical_application}
\input{41_numerical_scheme}
\input{42_results}

\input{43_taylor_green}

\input{44_numerical_scheme}
\input{45_results}

\input{50_conclusions}

\section*{Acknowledgments}
This work has been partially supported by the Swedish Research Council, under grant 2018-02438, and
the Ministry of Science, Innovation, and Universities, under research project RTI2018-096791-B-C21,
\textit{Hydrodynamics of Motion Damping Devices for Floating Offshore Wind Turbines (FOWT-DAMP)}. P.E. Merino is supported during the completion of his Ph.D. thesis by MEyFP grant FPU17/05433 and thanks MEyFP for its support.

\section*{Data availability}
The data that support the findings of this study are available from the corresponding author upon reasonable request.
\appendix

\input{60_boundaries}
\input{62_hamiltonian}
\input{61_delta}

%\section*{Bibliography}
\bibliography{bib}

\end{document}

%% file: 01_title.tex
\title{%[Time integration and energy conservation in SPH]{%
	The role of time integration in energy conservation in
	Smoothed Particle Hydrodynamics fluid dynamics simulations}
%
%	Smoothed Particle Hydrodynamics energy balance for fluid dynamics}
%
\author[uppsala]{Jose Luis Cercos-Pita\corref{corresponding}\fnref{noteJLC}}
%\affiliation{%
\address[uppsala]{%
	Hedenstierna laboratory, Surgical Sciences Department,\\
	Uppsala Universitet, 75185 Uppsala, Sweden}
\cortext[corresponding]{Corresponding author}
\ead{jl.cercos@upm.es}
\fntext[noteJLC]{ORCID: 0000-0002-3187-4048}


\author[masai]{Pablo Eleazar Merino-Alonso\fnref{notePEM}}
\address[masai]{%
	M2ASAI Research Group,\\
	ETS Ingenieros Navales, Universidad Polit\'ecnica de Madrid,\\
	28040 Madrid, Spain
}

\fntext[notePEM]{ORCID:  0000-0002-2630-3590 }

\author[cehinav]{Javier Calderon-Sanchez\fnref{noteJC}}

\fntext[noteJC]{ORCID: 0000-0003-0636-8853}

\author[cehinav]{Daniel Duque\fnref{noteDD}}

\fntext[noteDD]{ORCID: 0000-0002-2248-5630}

\address[cehinav]{%
CEHINAV Research Group,\\
ETS Ingenieros Navales, Universidad Polit\'ecnica de Madrid,\\
28040 Madrid, Spain
}

%% file: 02_abstract.tex
\begin{abstract}
The choice of a time integration scheme is a crucial aspect of any transient fluid simulation, and Smoothed-Particle Hydrodynamics (SPH) is no exception. 
The influence of the time integration scheme on energy balance is here addressed.
To do so, explicit expressions allowing to compute the deviations from the energy balance, induced by the time integration scheme, are provided. These expressions, computed \textit{a posteriori}, are valid for different integration methods.
%
%A new treatment of the energy balance is developed in this work, which explicitly includes a new term which is shown to be broadly related with tensile instability.
%
%In addition to that, the role of the time integration scheme in the energy balance is addressed.
%
Besides, a new formulation that improves energy conservation by enhancing stability, based on an implicit integration scheme, is proposed.
Such formulation is tested with the simulation of a two-dimensional non-viscous impact of two jets, with no artificial dissipation terms.
To the best of our knowledge, this is the first stable simulation of a non-dissipative system with a weakly-compressible SPH method.
A viscous case, the Taylor-Green vortex, has also been simulated. Results show that an implicit time integration scheme also behaves better in a viscous context.
\end{abstract}
%
%\begin{keyword}
%SPH, Energy balance, Tensile instability, Time integration, Jet impact
%\end{keyword}
\begin{keyword}
	stability, time integration scheme, energy balance, SPH
\end{keyword}

%% file: 10_intro.tex
\section{Introduction}
\label{s:intro}
Smoothed-Particle Hydrodynamics (SPH) is a meshfree numerical method in which continuum media are discretized as a set of particles, which move in a Lagrangian manner \cite{monaghan_arfm_2012}.
There is no doubt that its meshless nature is the feature which has drawn more attention to the model, that is indeed well suited to problems dominated by complex geometries, such as simulations involving free surface flow, or flows driven by large boundary displacements.

In addition to that, SPH is built starting from a relatively simple formulation that can be applied to a wide variety of physical phenomena.
Indeed, even though the model was initially developed in astrophysics \citep{lucy1977, gingold1977}, it quickly spread to other disciplines, including free surface flows \citep{mon1994b}, solid mechanics \citep{rabczuk2003, bonet2004} and geomaterial mechanics \citep{bui2008}.

An interesting feature of SPH which has been traditionally considered one of its main benefits is the conservation of both momentum and energy.
The claim that the method features exact energy conservation has been made several times in the past \citep{monaghan1983, couchman1995, vaughan2008}, although literature may also be found (e.g. \citep{price2008}) where such conservation is shown to be linked to the accuracy of the time integration algorithm.
Therefore, a clear line of investigation to improve the stability of the model comes from the analysis of the time integration scheme. Previous research on this topic has already shown promising results \citep{dyka1997, rook2007, colagrossi2010, blanc2012, mabssout2013}.

Indeed SPH has been widely criticized for characteristic instabilities, particularly in its weakly-compressible SPH (WC-SPH) incarnation, which is by far the most popular one.
%
%Researchers have conducted a number of studies about SPH stability, blaming a plethora of aspects of the model, while correspondingly providing a number of alternative formulations.
%
One of the most aggressive solutions to avoid stability issues of WC-SPH is to implement it within a rigorous incompressible formulation, leading to Incompressible-SPH (I-SPH) \citep{Cummins_Rudman_JCP1999}, a method that inherits some good stability features of more conventional CFD methods \citep{Xu20096703}, at the expense of increased algorithm complexity.

Within the WC-SPH formulation, and probably motivated by the excellent energy conservation properties, some authors attributed the instabilities to spurious zero-energy modes \citep{vignjevic2000, vidal2007}.
%
%These investigations have been losing favor against the closely-related line that focuses on tensile instability \citep{mon2000b, Swegle+etal:1995, dehnen_aly_wendland_2012}.
%
Lately, the focus has been set on tensile instability \citep{mon2000b, Swegle+etal:1995, dehnen_aly_wendland_2012}.
The first formulation designed to mitigate the pernicious effects of this instability was X-SPH, in which the velocity field is smoothed at each particle using information from its neighbors \citep{mon2000b}.
Along this line, some authors \citep{Swegle+etal:1995} analyzed the convolution kernel, culminating in the work by \citet{dehnen_aly_wendland_2012}, where it was demonstrated that Wendland kernels benefit particle packing.

Another methodology to deal with the tensile instability which is gaining popularity is the Particle Shifting Technique (PST), in which the particles' positions are slightly modified at the end of each time step in order to preserve particle packing \citep{Xu20096703,Lind20121499,vacondio2013variable,Colagrossi_CPC_2012,oger2016sph, sun_2019}.
%
%One year before, the same principle was also applied to the correction of initial particle positions.\cite{Colagrossi_CPC_2012}.
%
%PST was introduced for I-SPH by \citet{Xu20096703}, improved later, \citep{Lind20121499}
%
%and extended to WC-SPH. \cite{vacondio2013variable}
%
%One year before, the same principle was also applied to the correction of initial particle positions.\cite{Colagrossi_CPC_2012}.
%
%In the last decade, improved PST and analogously related techniques have been developed to deal satisfactorily with instabilities\cite{oger2016sph, sun_2019}.
%
Some authors who have dealt with tensile instabilities are moving to the so-called Total Lagrangian formalism \citep{vignjevic2006}, specially in solid dynamics \citep{bonet2002, zhang2014, islam2019}.
%
%In Total Lagrangian SPH, the convolution kernel is evaluated according to the initial particle positions in order to avoid instabilities.
%

A different line of investigation to improve stability has been the application of extra energy dissipation terms, 
the most straightforward one being artificial viscosity \citep{monaghan_arfm_2012}.
However, research quickly targeted mass conservation as well, by means of Shepard filtering.
%However, research quickly targeted mass conservation as well, first addressing the possibility of smoothing the density field in a similar fashion to what X-SPH does for the velocity field ---  a technique named Shepard filtering.
%
Afterwards, the addition of dissipation terms to the mass conservation equation has been investigated, resulting in the $\delta$-SPH \cite{AntuonoCPC2012}, and Riemann solvers-based schemes \cite{vila1999}.
%
%Later on, the addition of dissipation terms to the mass conservation equation has been investigated, resulting in two main different branches: $\delta$-SPH \citep{AntuonoCPC2012}, and Riemann solvers \citep{vila1999}.
%
The relation between both formulations has been addressed in the past \cite{CercosPita2016, green2019smoothed}.

In many of the research targeted at intrinsic instabilities just described, it is not yet clear how novel algorithms and formulations may affect the conservation properties of the method.
It is not unreasonable to suggest that these studies have been circumvented due to the fact that the SPH community has traditionally regarded SPH as an exact energy conservation model, and the efforts have been accordingly directed towards solutions to known drawbacks of the method.
%
% When you refer to a work (e.g. In <...> the authors blablabla), you should use citet
%
For instance, an energy analysis of the $\delta$-SPH term was conducted in \citet{antuono2015energy}, demonstrating that it is intrinsically dissipating energy far from the boundaries.
However, such energy dissipation is presented as a pernicious side effect of the model, a point which is not obvious, as discussed below.
Similarly, \citet{Green_Peiro_2018} examine energy conservation and partition between kinetic, potential, and compressible energies at the post-processing stage, in order to assess different models, for a long-duration simulation.

In \citet{CercosPita_etal_CMAME_2017_SPH_ENERGY}, violations of exact energy conservation were formally demonstrated for the first time.
In such work, fluid extensions are considered, and extra energy terms are shown to appear due to interactions with the boundary.
The investigation was also extended to other boundary formulations \citep{cercospita_thesis_2016}.

Surprisingly, although both the influence of the time integration scheme in the stability and the benefits of eventual extra energy dissipation have been already demonstrated, the role of the time integration scheme in the energy conservation has not been addressed in the literature yet.
This paper is therefore devoted to this topic.
%introducing the role of the time integration scheme in the SPH energy balance.
%
For the sake of simplicity, we focus on the WC-SPH formulation for non-viscous fluids.

In order to analyze energy balance, spatial and time discretization are independently considered: the former will be presented in Section \ref{s:spatial} and the latter in Section \ref{s:time}. They are combined afterwards in a total energy balance in Section \ref{ss:energy_balance}. In Section \ref{ss:implicit_time_step} an implicit time integration scheme is proposed in order to improve energy conservation.
Then, numerical experiments are carried out in Section \ref{s:application}, in order to support the theoretical findings.
One of the simulations presented in this Section would be the first stable simulation of a non-dissipative system with a weakly-compressible SPH method (to the best of our knowledge.)
The other is an application of this methodology to a viscous benchmark case.
%
%By carefully addressing the various ways in which energy may fail to remain constant, a final, stable simulation is here described.
Finally, conclusions are presented in Section \ref{s:conclusions}.

%% file: 20_spatial_discretization.tex
\section{Spatial discretization}
\label{s:spatial}
This section deals chiefly with spatial discretization. The SPH governing equations are
introduced in Section \ref{ss:gov_equations}. Power balance is discussed in Section \ref{ss:power}, where
contributions to energy variation are identified and separated.

%% file: 21_sph.tex
\subsection{SPH numerical model}
\label{ss:gov_equations}
Herein we focus on weakly compressible flows, even though similar analyses can be carried out for incompressible flows, or even for different physical phenomena.
Hence, the governing equation for the evolution of density is the conservation equation:
\begin{equation}
\label{eq:gov_equations:mass_cons}
\SPH{\frac{\D \rho}{\D t}}_i(t) = -\rho_i(t) \SPH{\Div{\bs{u}}}_i(t) %+ \SPH{\frac{\D \rho}{\D t}}^\delta_i(t),
\end{equation}
where $\SPH{\cdot}$ denotes SPH operators, and abusing the notation, any magnitude resulting from the application of the SPH methodology.
In the equation above $\rho_i$ is the density of an arbitrary $i$-th particle, and $\bs{u}_i$, its velocity.
%The last term represents terms introduced by a possible $\delta$-SPH treatment. \citep{AntuonoCPC2012, CercosPita2016}

The evolution of the velocity field is a discrete version of the Navier-Stokes momentum equation,
\begin{equation}
\label{eq:gov_equations:mom_cons}
	\SPH{\frac{\D \bs{u}}{\D t}}_i(t) =
		- \frac{\SPH{\Grad{p}}_i(t)}{\rho_i(t)}
		+ \frac{\mu}{\rho_i(t)} \SPH{\Lap{\bs{u}}}_i(t)	
		+ \bs{g} 
		- k \frac{p_i \SPH{\Grad{\gamma}}_i(t)}{\rho_i(t)}
		,
\end{equation}
where $p_i$ is the pressure, $\mu$ the viscosity coefficient, $\bs{g}$ the acceleration due to external forces, \added{and $\gamma$ the Shepard renormalization factor (see, for instance \cite{calderon_etal_cpc_2018})}. The extra term with the coefficient $k$, which appears in this discrete version and is absent in the continuum, is explained below.

In WC-SPH, these equations are closed by an equation of state (EOS) relating pressure and density:
\begin{equation}
\label{eq:gov_equations:eos}
 p_i(t) = p_0 + c^2_0 \, (\rho_i(t) - \rho_0),
\end{equation}
where $p_0$ is the background pressure, $\rho_0$ is the reference density, and $c_0$ the speed of sound in the fluid. %
The latter is customarily set to a value high enough that the fluid behaves almost as if incompressible.

Incidentally, it may be highlighted that every single SPH related operator can be split in 2 terms,
\begin{equation}
\label{eq:gov_equations:sph_split}
\SPH{\cdot} = \SPH{\cdot}^\Omega + \SPH{\cdot}^{\partial\Omega},
\end{equation}
i.e., all SPH operators can be split in the effect of the interactions with other fluid particles and the interactions with the boundary.
The latter may adopt a number of forms \cite{cercospita_thesis_2016}.
For the sake of simplicity, only the first, ``bulk'', part of the operators is considered in the main body of this article, and boundary effects are deferred to \ref{ss:boundaries}.

Regarding the viscous term $\SPH{\Lap{\bs{u}}}$, it may take several forms \citep{mon1992,morris1997b}, and may even include bulk viscosity \cite{PhysRevE.101.013302}.
As with boundary effects, we leave this term undefined, for the sake of simplicity and generality.

The momentum equation \eqref{eq:gov_equations:mom_cons} differs from the traditional SPH formalism (as e.g. in the works of \citet{monaghan_arfm_2012} and \citet{violeau_2012_book}), by an extra term involving \(\SPH{\Grad{\gamma}}^{\Omega}\), whose form is:

\begin{equation}
	\label{eq:gov_equations:grad_gamma}
	\SPH{\Grad{\gamma}}^\Omega_i(t) = \sum_{j \in \Omega}  \Grad W_{ij} \frac{m_j}{\rho_j(t)} ,
\end{equation}
where $m_j$ the mass of an arbitrary $j$-th particle, $W_{ij} := W(|\bs{r}_i - \bs{r}_j|)$ is the value of kernel function for particles $i$ and $j$,
and $\Grad W_{ij} := \Grad W (|\bs{r}_i - \bs{r}_j|) $.
This term consistently vanishes in the continuum, but not in the discrete formulation of the equations, except for highly ordered particle distributions, such as those often set up at the beginning of a simulation. 
This form of momentum equation is introduced by \citet{Colagrossi_CPC_2012}, where the consequences of such a term are thoroughly discussed.

Equations \eqref{eq:gov_equations:mass_cons} and \eqref{eq:gov_equations:mom_cons}  feature discrete divergence and gradient operators. For these, anti-symmetric SPH operators $\SPH{\Div{\bs{u}}}$ and $\SPH{\Grad{p}}$ may be considered, of the form:
\begin{align}
	\label{eq:gov_equations:div_u}
		\SPH{\Div{\bs{u}}}_i(t)  &= \sum_{j \in \Omega}
		\left[
		\bs{u}_j(t) - \bs{u}_i(t)
		\right] \cdot \Grad W_{ij} \frac{m_j}{\rho_j(t)}  ,
	\\
	\label{eq:gov_equations:grad_p}
		\SPH{\Grad{p}}_i(t) & = \sum_{j \in \Omega}
		\left[
		p_j(t) - p_i(t)
		\right] \Grad W_{ij} \frac{m_j}{\rho_j(t)} ,
\end{align}
%
% O(h) error if we consider the boudnaries, O(h^2) if we don't
%
which are consistent with an $O(h^2)$ error.

If the value $k $ is set to $2$ in Eq. \eqref{eq:gov_equations:mom_cons}, the resulting equation is effectively equivalent to using the popular symmetric pressure gradient operator \citep{monaghan_arfm_2012,violeau_2012_book}
\begin{equation}
	\label{eq:gov_equations:grad_p_sym}
	\SPH{\Grad{p}}^\text{sym}_i(t)  = \sum_{j \in \Omega}
	\left[
	p_j(t) + p_i(t)
	\right]
	\Grad W_{ij} \frac{m_j}{\rho_j(t)} .
\end{equation}

This formulation has been already considered by \citet{Colagrossi_CPC_2012} in the context of setting an optimal initial particle distribution.
%Setting \added{$k = 2$} in Eq. \eqref{eq:gov_equations:mom_cons} results in the same overall effect as using the popular symmetric pressure gradient operator \citep{monaghan_arfm_2012,violeau_2012_book}.
%
A symmetric pressure gradient operator grants immediate linear and angular momentum conservation, as well as power conservation \citep{mon2005}.
However, we prefer to keep this term and the anti-symmetric gradient operator as two distinct terms, since the interplay between the two will be crucial, as further discussed below.
%
%The possibility of considering both symmetric and anti-symmetric SPH operators has been justified in the past by the fact that the %$\SPH{\Grad{\gamma}}$ term consistently vanishes in the continuum, but not anymore in the discrete.
%
The symmetry properties of these operators also play a main role in multiphase flow simulations and free-surface modelling, as discussed by \citet{Colagrossi2009}.

Finally, even if a linear EOS is suggested in Eq. \eqref{eq:gov_equations:eos}, other expressions have been traditionally accepted in SPH, provided this more general relationship holds:
\begin{equation}
\label{eq:gov_equations:dpdt}
\frac{\D p_i}{\D t}(t) = c^2_0 \, \frac{\D \rho_i}{\D t}(t).
\end{equation}

%% file: 22_sph_energy_balance.tex
\subsection{SPH Power balance}
\label{ss:power}
Analyses of power balance in SPH have been carried out several times in the past, e.g.: \citep{antuono2015energy, cercospita_thesis_2016, CercosPita_etal_CMAME_2017_SPH_ENERGY}.
Every time that the balance is revisited, new terms are included, and a better arrangement of the existing ones is proposed.
In this work, power balance is critical, so its development is carried out from the basics in this section.
Incidentally, it should be highlighted that in this work ``power balance'' refers to the instantaneous balance, which only depends on the spatial SPH operators, while ``energy balance'' involves the time integration scheme.

Thus, temporarily accepting the acceleration term computed in Eq. \eqref{eq:gov_equations:mom_cons} as the actual rate of variation of the particle velocity, the following kinetic power term can be defined,
\begin{equation}
\label{eq:power:kinetic_sph}
\SPH{P\kin}(t) := \sum_{i \in \Omega} m_i \, \bs{u}_i(t) \cdot \SPH{\frac{\D \bs{u}}{\D t}}_i(t),
\end{equation}
which is indeed the discrete version of the continuous kinetic power.
This kinetic power can be conveniently split in several terms applying Eqs. \eqref{eq:gov_equations:mom_cons} and \eqref{eq:gov_equations:sph_split},
\begin{align}
\label{eq:power:balance}
	\SPH{P\kin}(t) &= \SPH{P_{\Grad{p}}}(t) + \SPH{P_{\mu}}(t) - P_p(t) +
	k \SPH{P_{\Grad{\gamma}}}(t) ,
\\
\label{eq:power:grad_p}
	\SPH{P_{\Grad{p}}} &= - \sum_{i \in \Omega}  \bs{u}_i(t) \cdot \SPH{\Grad{p}}_i(t) \frac{m_i}{\rho_i(t)},
\\
\label{eq:power:viscous}
	\SPH{P_{\mu}} &= \mu \sum_{i \in \Omega}\bs{u}_i(t) \cdot \SPH{\Lap{\bs{u}}}_i(t) \frac{m_i}{\rho_i(t)} ,
\\
\label{eq:power:potential}
	P_p(t) &= -\sum_{i \in \Omega}  m_i \bs{u}_i(t) \cdot \bs{g},
\\
\label{eq:power:gradgamma}
	\SPH{P_{\Grad{\gamma}}} &= - k \sum_{i \in \Omega} p_i(t) \bs{u}_i(t) \cdot \SPH{\Grad{\gamma}}_i(t)
	\frac{m_i }{\rho_i(t)} .
\end{align}
The first three terms above, Eqs. (\ref{eq:power:grad_p}-\ref{eq:power:potential}), are just discrete versions of well known terms in fluid dynamics. To be precise: $\SPH{P_{\Grad{p}}}$ corresponds to the work due to pressure, $\SPH{P_{\mu}}$, to the power due to the viscous forces, and $P_p$ to power from conservative external forces.

The extra term \eqref{eq:power:gradgamma} is due to the $\SPH{\Grad{\gamma}}$ operator, and obviously vanishes if $k = 0$.
Paradoxically, the vanishing of this term is not beneficial for the numerical scheme \citep{mon2005}.
This can be easily checked out by rearranging the work due to pressure, $\SPH{P_{\Grad{p}}}$, as the power due to compressibility, as it is usually done in fluid dynamics.
To this end, the following identity, which is often used in order to prove that the anti-symmetric divergence \eqref{eq:gov_equations:div_u} and the symmetric gradient \eqref{eq:gov_equations:grad_p_sym} operators are skew-adjoint \citep{mayrhofer_etal_cpc_2013,antuono2015energy} can be invoked:
\begin{equation}
\label{eq:power:sym_asym_model}
-\sum_{i \in \Omega} 
\bs{u}_i(t) \cdot \SPH{\Grad{p}}^\text{sym}_i(t)
\frac{m_i }{\rho_i(t)}
=
\sum_{i \in \Omega}
 p_i(t) 
	\SPH{\Div{\bs{u}}}_i(t) 
\frac{m_i  }{\rho_i(t)  } .
\end{equation}

See \ref{s:hamiltonian} for a straightforward calculation that shows that any pair of gradient and divergence operators that satisfy this identity will lead to the conservation of mechanical energy.

Eq.~\eqref{eq:power:sym_asym_model} can be written in terms of the anti-symmetric gradient \eqref{eq:gov_equations:grad_gamma} as
% Section \ref{ss:gov_equations},
%
\begin{equation}
\label{eq:power:grad_p_to_divu}
-\sum_{i \in \Omega}
\bs{u}_i(t) \cdot \left[
	\SPH{\Grad{p}}_i(t) + 2 p_i(t) \SPH{\Grad{\gamma}}_i(t)
\right]
\frac{m_i }{\rho_i(t)}
 =
\sum_{i \in \Omega}
 p_i(t)
\SPH{\Div{\bs{u}}}_i(t)
\frac{m_i }{\rho_i(t)} .
\end{equation}
Using \added{the identity of Eq.~\eqref{eq:power:grad_p_to_divu} in the work due to pressure,} Eq.~\eqref{eq:power:grad_p}, and applying mass conservation, Eq. \eqref{eq:gov_equations:mass_cons}, the work due to pressure can be subsequently rearranged as
\begin{align}
\label{eq:power:grad_p:balance}
	\SPH{P_{\Grad{p}}}(t) &=  - \SPH{P_c}(t) - 2 \SPH{P_{\Grad{\gamma}}}(t)
\\
\label{eq:power:compressible_sph}
	\SPH{P_\mathrm{c}}(t) & = \sum_{i \in \Omega}
	\frac{p_i(t)}{\rho_i(t)} \SPH{\frac{\D \rho}{\D t}}_i \!\! (t)\,
	\frac{m_i }{\rho_i(t)} 	,
\end{align}
where $\SPH{P_\mathrm{c}}$ is the power due to the compressibility.

Finally, all these terms may be introduced in the power balance, Eq. \eqref{eq:power:grad_p:balance}, which can then be rewritten
\begin{equation}
\label{eq:power:final_balance}
	\SPH{P\kin}(t) + P_p(t) + \SPH{P_c}(t) 	=
\\
\SPH{P_{\mu}}(t)  +    \left( k - 2 \right) \SPH{P_{\Grad{\gamma}}}(t)
\end{equation}
The term at the right hand side is an extra term, whose presence causes a lack of energy conservation
(in addition to the term due to viscosity, which has a physical basis).
Hence, it is possible to automatically ensure exact power balance if a symmetric pressure gradient operator ($k = 2$) is chosen.
In other words: the symmetric pressure gradient operator enables the model to non-physically exchange kinetic and compressibility energies, for the sake of total power conservation.
\added{%
Consistently, the same calculation for the conservation of mechanical energy referred to above will arrive at an expression that is not constant in general, unless $k=2$. }
Such power conservation feature of the symmetric pressure gradient operator was already discussed by \citet{mon2005}.

A number of works \added{(e.g. \citep{Bonet_Lok_1999}, \cite{Colagrossi2009})} have targeted the SPH operator $\SPH{\Grad{\gamma}}$. 
\added{Some authors, \cite{sun_2018} \cite{Lind20121499},  proposed to diminish the value of such term by different means.}
However, its energy balance term, $\SPH{P_{\Grad{\gamma}}}$, has not been addressed in the past.
%
%This is further discussed below, in Section \ref{ss:particle_shifting}.

\replaced{The effect of boundaries is avoided in the present work, as this introduces additional terms to the equations, as discussed in}{We stress that this discussion has avoided the effects of boundaries, which introduce additional terms to the equations, as discussed in} \ref{ss:boundaries}. Also, the SPH model is a simple one: other formulations such as $\delta$-SPH would \deleted{also} add additional terms, see \ref{ss:deltaSPH}.

%% file: 30_time_discretization.tex
\section{Time discretization}
\label{s:time}

After time discretization, the semi-discrete magnitudes, $f_i(t), \sph{f}_i(t)$ (discrete in space but continuous in time) lead to fully discretized magnitudes, $f_i^n, \sph{f}_i^n$,
where the superindex $n$ indicates evaluation at the n-th time step.
This may be achieved by a number of different time integration schemes.
In particular, several ones have been implemented in the most popular codes, e.g. improved Euler \citep{CercosPita2015}, Verlet \citep{Crespo2015204}, and second-order Runge-Kutta \citep{herault_bilotta_dalrymple_jhr10}.
The benefits and drawbacks of some of these integration schemes have been addressed in the past, mainly from the point of view of stability and time step size \citep{dyka1997, rook2007, colagrossi2010, mabssout2013}.
%

%Here, a general Euler notation is first established in \ref{ss:euler}. The time variation of the kinetic energy is studied in \ref{ss:energy_balance}. A possible implicit scheme is proposed in \ref{ss:implicit_time_step}, which could improve energy conservation. In Section \ref{ss:particle_shifting}, possible PST strategies are discussed.
The fact that the integration in time is numerically performed, naturally implies an error. The presence of such a deviation term has already been introduced in the past (see, among others, \citet{mon2005} and \citet{price2008}), although its analysis has always been left aside.

The purpose of this section is to provide analytical expressions to evaluate these residuals in the most general case.

%% file: 31_time_scheme.tex
\subsection{Equivalent Euler time scheme}
\label{ss:euler}
In order to analyze a general integration scheme, the actual variation rates are computed \textit{a posteriori}. These variation rates are defined, within a time step, as
%
% Former eqs 21 and 22 were the same one
%
\begin{equation}
\label{eq:varrates}
\vrate{f}{i}{n} := \dfrac{f_i^{n+1}- f_i^n}{\Delta t},
\end{equation}
where $\Delta t = t^{n+1}- t^n$ is the time step.

According to this, we define 
\begin{align}
\label{eq:gov_equations:drhodt_euler}
	\vrate{\rho}{i}{n} &:=
	\frac{\rho_{i}^{n + 1} - \rho_{i}^n}{\Delta t},
 \\
\label{eq:gov_equations:dudt_euler}
	\vrate{\bs{u}}{i}{n} &:=
	\frac{\bs{u}_{i}^{n + 1} - \bs{u}_{i}^n}{\Delta t}.
\end{align}
\added{Let us remark that the variation rates defined by \eqref{eq:varrates}, \eqref{eq:gov_equations:drhodt_euler} and \eqref{eq:gov_equations:dudt_euler} can only be computed \textit{a posteriori}, that is, after the flow fields at time $n+1$ are obtained by the actual time integration algorithm. Let us also point out that these variation rates serve as an analysis tool to asses the effect of the time integration scheme, but are not involved in the time integration itself.}

With this in mind, an ``equivalent Euler integration scheme'' can be defined \textit{a posteriori}. This ``equivalent scheme'' gives, within the n-th time step and using the rates defined by \eqref{eq:varrates}, the same result as the considered time integration scheme using the actual variation rates, given by the SPH equations, and applying the corresponding algorithm.

%% file: 32_total_energy_balance.tex
\subsection{SPH total energy balance}
\label{ss:energy_balance}
Considering the total kinetic energy of the discrete system, at any time step,
\begin{equation}
E\kin^n = \frac{1}{2} \sum_{i \in \Omega} m_i \, \left\vert \bs{u}_i^n \right\vert^2,
\end{equation}
we can write the rate of total kinetic energy variation during consecutive time steps as
\begin{equation}
\label{eq:energy:kinetic}
P\kin^{n^{\ast}} :=
\vrateni{E\kin}{n} = 
\frac{E\kin^{n + 1} - E\kin^{n}}{\Delta t}. 
\end{equation}

Developing this term one gets 
\begin{equation}
P\kin^{n^\ast} = \dsty{\sum_{i \in \Omega}} m_{i}
\left( \dfrac{\bs{u}_{i}^{n+1} + \bs{u}_{i}^n}{2} \right)
\cdot
\left( \dfrac{\bs{u}_{i}^{n+1} - \bs{u}_{i}^n}{\Delta t}\right).
\end{equation}
Substituting $\bs{u}_{i}^{n+1}$ by its value, according to the equivalent Euler scheme, gives
\begin{equation}
\label{eq:Pkrate}
P\kin^{n^\ast} = \dsty{\sum_{i\in \Omega}} m_{i} 
\left[
\bs{u}_i^n + \dfrac{\Delta t}{2} \vrate{\bs{u}}{i}{n}
\right]
\cdot
\vrate{\bs{u}}{i}{n}.
\end{equation}
Let us now compute the deviation of this power away from that of Eq. \eqref{eq:power:kinetic_sph}, caused by the errors in the velocity field due to the time integration scheme:
\begin{equation}
\label{eq:Pkerror}
P_{\D \bs{u} / \D t}(t) := P\kin^{n^\ast} - \sph{P\kin}(t),
\end{equation}
for any time $t \in (t_n, t_{n+1})$. 
Substituting \eqref{eq:Pkrate} and \eqref{eq:power:kinetic_sph} in \eqref{eq:Pkerror} we get
\begin{equation}
\label{eq:powerres}
P_{\D \bs{u} / \D t}(t) = 
\dsty{\sum_{i \in \Omega}} m_{i} 
\left[
\left(
\bs{u}_i^n + \dfrac{\Delta t}{2} \vrate{\bs{u}}{i}{n}
\right)
\cdot
\vrate{\bs{u}}{i}{n}
- 
%\right.\notag\\&\left.
\bs{u}_{i}(t) \cdot \sph{\dfrac{\D \textbf{u}}{\D t}}_i(t) 
\right].
\end{equation}

On the other hand, the velocities computed with the fully discretized SPH scheme do not agree with the ones resulting from solving the semidiscrete scheme, as the time integrations involve an error.

%%
%\begin{equation}
%\textbf{u}_i^{n} = \textbf{u}_i^0 + \Delta t \dsty{\sum_{m=0}^{n-1}} \vrate{\bs{u}}{i}{m}.
%\end{equation}
%%
%Inserting this expression in \eqref{eq:powerres} we get
%%
%\begin{align}
%\label{eq:power-error-1}
%P_{\D \bs{u} / \D t}(t) = 
%\dsty{\sum_{i \in \Omega}} m_{i} \Bigg\{
%&\left[ \bs{u}_i^0 + \Delta t \left(\dsty{\sum_{m=0}^{n-1}} \vrate{\bs{u}}{i}{m} + 
%\dfrac{1}{2}   \vrate{\bs{u}}{i}{n}
%\right) \right]
%\cdot \vrate{\bs{u}}{i}{n}
%%\right. 
%\notag \\ 
%%\left.
%- &\bs{u}_{i}(t) \cdot \sph{\dfrac{\D \textbf{u}}{\D t}}_i(t)
%\Bigg\}.
%\end{align}

Furthermore, the SPH approximation of the derivative, for $t \in (t_n, t_{n+1})$ can be split into two terms:
\begin{equation}
\label{eq:deltapow}
\sph{\dfrac{\D \bs{u}}{\D t}}_i(t) = \vrate{\bs{u}}{i}{n} - \bs{\delta}^n(t)
\end{equation}
where $\bs{\delta}^n(t)$ is the vector difference, at time $t \in (t_n, t_{n+1})$, between the SPH and the Equivalent Euler \textit{a posteriori} variation rates. According to this, \eqref{eq:powerres} can be rewritten as
%
%\colblue{
\begin{align}
\label{eq:power-error-2}
P_{\D \bs{u} / \D t}(t) = 
&\dsty{\sum_{i \in \Omega}} m_{i} 
\left[
\left(
\bs{u}_i^n + \dfrac{\Delta t}{2} \vrate{\bs{u}}{i}{n} - \bs{u}_i(t)
\right)
\cdot
\vrate{\bs{u}}{i}{n}
\right] + 
\notag \\
+ &\dsty{\sum_{i \in \Omega} }m_{i} \textbf{u}_i(t) \cdot \bs{\delta}^n(t).
%
%
%
%P_{\D \bs{u} / \D t}(t) = 
%&\dsty{\sum_{i \in \Omega} } m_{i} 
%\left\{
%\left[
%\bs{u}_i^0 + \Delta t \left(\dsty{ \sum_{m=0}^{n-1} } \vrate{\bs{u}}{i}{m} 
%%+ \right.\right.\right.
%%\notag \\ 
%%&\left.\left.\left.
%+ \dfrac{1}{2} \vrate{\bs{u}}{i}{n}
%\right) \right] 
%- \bs{u}_{i}(t)\right\}
%\cdot
%\vrate{\bs{u}}{i}{n}
%\notag \\
%&\dsty{ + \sum_{i \in \Omega} }m_{i} \textbf{u}_i(t) \cdot \bs{\delta}^n(t).
\end{align}
%}
%
The two terms in the sum above represent, respectively:
\begin{itemize}
\item The error due to approximating $\bs{u}_{i}(t_n)$ by $\bs{u}_i^n$, due to considering each $\vrate{\bs{u}}{i}{m}$ as the variation rate of $\bs{u}$ during the $(t_m, t_{m+1})$ time step.
\item The error committed in the variation rate, as expressed by \eqref{eq:deltapow}.
\end{itemize}

Let us now return to equation \eqref{eq:power:final_balance}.
It should be noticed that this energy balance % in Eq. \eqref{eq:power:final_balance}
is only valid if all the magnitudes involved are considered at the same time instant.

In order to evaluate the error using equation \eqref{eq:powerres}, we need to evaluate $\dsty{\SPH{\frac{\D \bs{u}}{\D t}}_i(t^{\ast})}$ at some $t^{\ast} \in (t_n,\, t_{n+1})$
This is therefore, computationally, a suboptimal approach.
On the other hand, this methodology can be applied to any time integration scheme.

In order to close the balance after applying Eq. \eqref{eq:power:final_balance}, similar error terms should be defined for the power due to the compressibility.
To that aim, let us examine the nature of equation \eqref{eq:Pkrate}. The variation rate of the kinetic Energy defined by that expression can be seen as an extended kinetic power, built from the discrete velocity field. In effect, that equation represents an exact evaluation of the kinetic power if we consider a linearly extended velocity field. Using the variation rate for the velocity defined by \eqref{eq:gov_equations:dudt_euler}, we define:
\begin{equation}
\label{eq:gov_equations:u_tau}
\overline{\bs{u}_i} (t_n + \tau) := \bs{u}_i^n + \tau \vrate{\bs{u}}{i}{n},
\end{equation}
\added{with $\tau \in [0, \Delta t]$.}
In effect, equation \eqref{eq:Pkrate} corresponds to the exact kinetic power of a set of particles, moving in the way described by the field \eqref{eq:gov_equations:u_tau}, evaluated at $t^{\ast} = t_n + \Delta t/2$. 

Therefore, a way of building an error term for the power due to compressibility, $P_{\D \rho / \D t}(t)$, is to assume linearly extended fluid fields evaluated at $t^{\ast} = t_n + \Delta t/2$. With that in mind, consider the following extended fields according to the variation rates defined by \eqref{eq:gov_equations:drhodt_euler} \eqref{eq:gov_equations:dudt_euler}:
\begin{align}
\label{eq:gov_equations:rho_tau}
	\overline{\rho_{i}}(t_n + \tau) &= \rho_{i}^n + \tau \vrate{\rho}{i}{n},
 \\
\label{eq:gov_equations:r_tau}
	\overline{\bs{r}_{i}}(t_n + \tau) &= \bs{r}_{i}^n + \tau \vrate{\bs{r}}{i}{n},
\end{align}
Therefore, defining
\begin{align}
\label{eq:energy:compressible_balance}
P_{\D \rho / \D t}(t^{\ast}) &:= P_\mathrm{c}^{n^{\ast}} - \SPH{P_\mathrm{c}}(t^{\ast}), 
\end{align}
and inserting our definition for $ P_\mathrm{c}^{n^{\ast}} $,
\begin{align}
\label{eq:power:compressible}
P_\mathrm{c}^{n^{\ast}} & =
\sum_{i \in \Omega}
\frac{\overline{p_i}(t^{\ast})}{\overline{\rho_i}(t^{\ast})}
\vrate{\rho}{i}{n}
\frac{m_i}{\overline{\rho_i}(t^{\ast})},
\end{align}
finally yields
\begin{align}
\label{eq:power:drhodt}
P_{\D \rho / \D t}(t^{\ast}) &=
\sum_{i \in \Omega} 
\left\{
\frac{\overline{p_i}(t^{\ast})}{\overline{\rho_i}(t^{\ast})}
\vrate{\rho}{i}{n}
\frac{m_i}{\overline{\rho_i}(t^{\ast})}
-
\frac{p_i(t^{\ast})}{\rho_i(t^{\ast})}
\SPH{\frac{\D \rho}{\D t}}_i(t^{\ast})
\frac{m_i}{\rho_i(t^{\ast})}
\right\}.
\end{align}
This allows to introduce a final balance:
\begin{equation}
\label{eq:energy:final_balance}
	 P\kin^{n^{\ast}} + P_p(t^{\ast}) + P_\mathrm{c}^{n^{\ast}} -
	 \SPH{P_{\mu}}(t^{\ast})    =
	 P_{\Delta t}(t^{\ast})
	+ \left( k - 2 \right) \SPH{P_{\Grad{\gamma}}}(t^{\ast}) ,
\end{equation}
where $P_{\Delta t}$ is a combined deviation term:
\begin{equation}
\label{eq:energy:p_deltat}
P_{\Delta t}(t^{\ast}) = P_{\D \bs{u} / \D t}(t^{\ast}) + P_{\D \rho / \D t}(t^{\ast}).
\end{equation}

Again, all terms at right hand side of Eq. \eqref{eq:energy:final_balance} have to be considered as extra terms that
cause a deviation from constant energy.
It is therefore patent that $P_{\Delta t}$ enters as an additional deviation term, whose expression depends on the particular time integration scheme.

%% file: 33_implicit_time_step.tex
\section{Implicit time step to improve the energy conservation}
\label{ss:implicit_time_step}
As discussed, in order to achieve exact energy conservation it is mandatory that the term $P_{\Delta t}(t^{\ast})$ vanishes.
%the terms $P_{\Delta t}(t_{n + 1/2})$ and $P_{\D \bs{r} / \D t}(t_{n + 1/2})$ vanish.
%
The most straightforward approach would consist on making the separate sub-terms, $ P_{\D \bs{u} / \D t}(t^{\ast})$, and $P_{\D \rho / \D t}(t^{\ast})$, identically vanish particle-wise for some $t^{\ast} \in (t_n, t_{n+1})$:
\begin{align}
\label{eq:gov_equations:dudt_implicit}
\vrate{\bs{u}}{i}{n} &= \SPH{\frac{\D \bs{u}}{\D t}}_i(t^{\ast}),
\\
\label{eq:gov_equations:drhodt_implicit}
\vrate{\rho}{i}{n} &= \SPH{\frac{\D \rho}{\D t}}_i(t^{\ast}) .
\end{align}

Diferent strategies may be used in order to satisfy \eqref{eq:gov_equations:dudt_implicit} and \eqref{eq:gov_equations:drhodt_implicit} or, at least, reduce the difference between the actual and the SPH variation rates.
In \cite{labudde-greenspan}, an interesting perspective, consisting on adapting the force computation accordingly to the time integration scheme, is proposed.
Another approach would be to act over the integrator. Along that line, implicit methods may be proposed. It is a well known fact  that implicit methots improve the stability of numerical integration of ODEs. Here, an implicit midpoint method is proposed
These methods, as will be discussed below, start from a guess for a mid-point position, which is iteratively refined until some condition (in our case, the previous equations) is satisfied within some tolerance.
On the other hand, larger time steps could be employed,
thereby perhaps compensating the extra computational cost per time step, since implicit methods are known to have better stability features \citep{leimkuhler_reich_2005}.
These aspects will be further discussed in the next section.

\subsection{An implicit midpoint method}
\label{sec:method}

Regarding the time discretization, several schemes have been considered in this work:
\begin{enumerate}
	\item Explicit Euler scheme: it is the simplest integration scheme, in which values of the derivative at some time step are used to evaluate values at the next step. The scheme is first order and not A-stable. \added{A-stability is a category of ODE solvers for which the solution is always stable, see \citep{suli_mayers_2003}.}
	\item Explicit Heun method: also known as improved \citep{suli_mayers_2003} or modified \citep{burden_2015} Euler's method, it is an explicit midpoint method. Simple predictor-corrector in which the explicit Euler method is used to obtain a prediction, which is then used in the correction. It is second order and not A-stable.
	\item Implicit midpoint method: a predictor-corrector scheme in which the mid point is guessed, and then iteratively refined. It is second order and A-stable.
\end{enumerate}

In the particular case of the implicit midpoint method, the time in which the fields are evaluated is $t^{\ast} = t_{n+1/2}$. In order to solve the implicit equation at every time step, a solution strategy must be defined, such as an iteration to find a fixed solution.
To this end, initial guesses $\lambda^0$ and $\mu^0$ for the velocity and density variation rates, respectively, are (we use superscripts as iteration indices):
\begin{align}
\bs{\lambda}_i^0 = \vrate{\bs{u}}{i}{n-1},
\\
\mu_i^0 = \vrate{\rho}{i}{n-1},
\end{align}
which start an iterative process. The values of the extended velocity and density fields at each iteration are noted by $\hat{\bs{u}}^m$ and $\hat{\rho}^m$, respectively. The SPH approximations to these fields at each iterations are noted $\SPH{\frac{\D \hat{\bs{u}}}{\D t}}_i$ and $\SPH{\frac{\D \hat{\rho}}{\D t}}_i$. With this in mind, one has
\begin{align}
\hat{\rho}^m_i(t_{n + 1/2}) & = \rho_i^n + \frac{\Delta t}{2} \mu_i^m,
\\
\mu_i^{m+1} & =
   f^m \SPH{\frac{\D \hat{\rho}}{\D t}}_i(t_{n + 1/2})
 + (1 - f^m) \mu_i^m
\\
 \hat{\bs{u}}^m_i(t_{n + 1/2}) & =
 \bs{u}_i^n +
 \frac{\Delta t}{2} \bs{\lambda}^m_i,
\\
  \bs{\lambda}^{m+1}_i &=
   f^m \SPH{\frac{\D \hat{\bs{u}}_j}{\D t}}_i(t_{n + 1/2})
 + (1 - f^m) \bs{\lambda}^m_i.
\end{align}
In the expressions above $f^m$ is the relaxation factor at iteration $m$, which has been set to
\begin{equation}
f^m = \frac{3}{4} \exp \left( \frac{
	-\left( m - \frac{M}{2} \right)^2
}{
	\sigma_f
}\right),
\end{equation}
where $M$ is the maximum number of iterations (either $10$ or $30$, in this practical application), and $\sigma_f$ a value tuned to get a small initial and final relaxation factor, $f^0 = f^M = 10^{-2}$.

%% file: 40_practical_application.tex
\section{Numerical verification}
\label{s:application}
In order to focus only on compressibility and kinetic energies, a practical application based on the frontal impact of two jets has been selected. Results are described next, in Section \ref{ss:normal_jets}.
However, legitimate concerns may arise regarding the role of the implicit time-integration scheme on a viscous simulation. Thus, a simulation of the Taylor-Green Vortex has also been carried out, with results in Section \ref{ss:taylor_green}.

\subsection{Frontal jet impact}
\label{ss:normal_jets}
This practical application has already been considered in order to assess energy conservation in SPH by \citet{marrone2015}.

\begin{figure}
	\centering
	\includegraphics[width=0.49\allwidth]{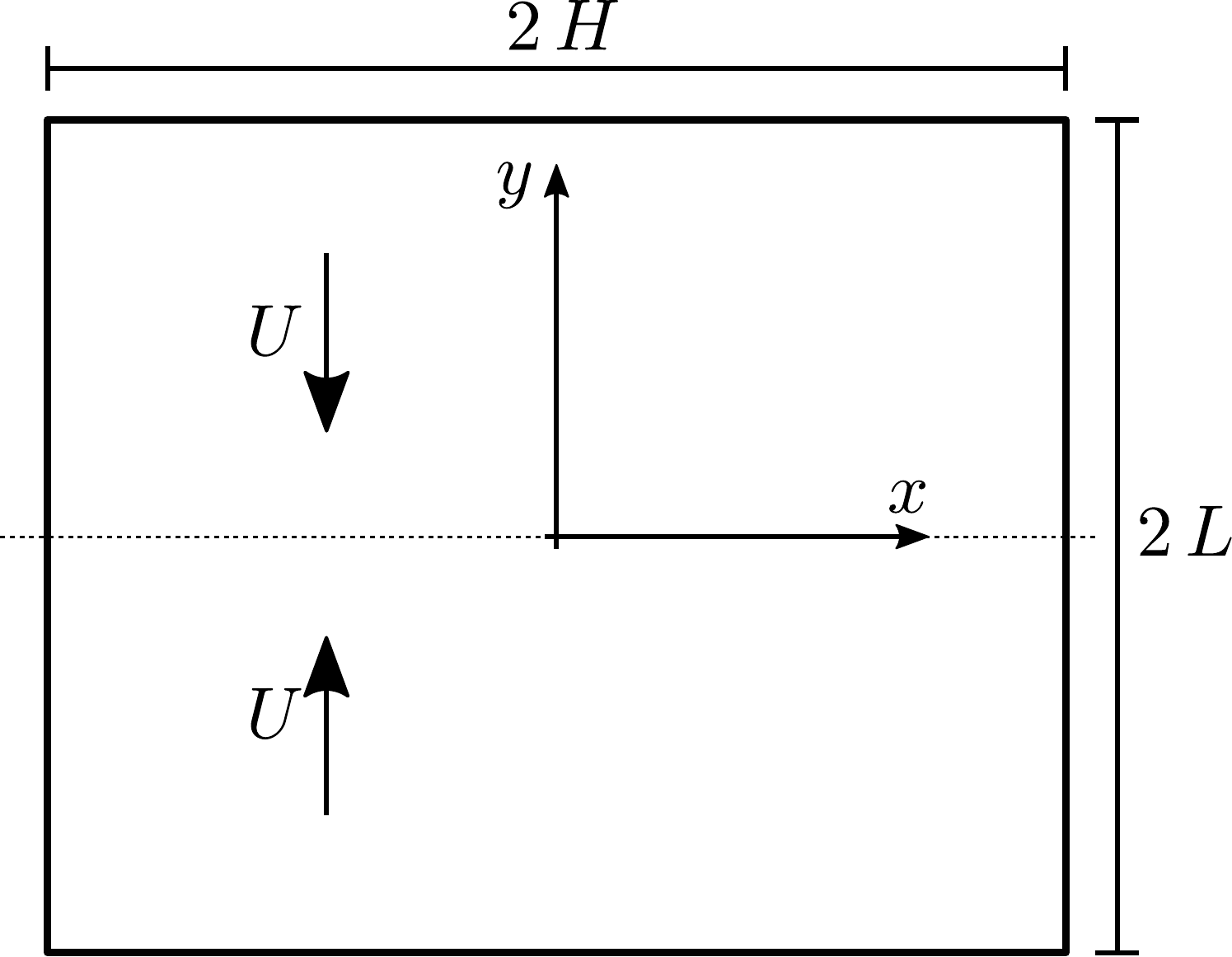}
	\caption{Schematic view of the initial condition for the water jets impingement simulation.}
	\label{fig:normal_impact:scheme}
\end{figure}

The initial condition is schematically depicted in Fig. \ref{fig:normal_impact:scheme}.
Two identical water jets of dimensions $2 H \times L$ each, with the same constant velocity $U$, but opposite directions, collide along the $y = 0$ plane at the initial time instant, $t = 0$.
Inviscid flow with a initially constant density field, $\rho = \rho_0$, is considered for both jets.
Also, no background pressure, $p_0 = 0$, or volumetric forces, $\bs{g} = 0$ are considered.
Thus, the whole phenomenon is characterized by the Mach number, $ \text{Ma} = U / c_0$.

When incompressible flow is considered, the impact causes a sudden loss
of a fraction of the initial energy dependent on the parameters chosen \cite{Szymczak_1994}.
Along this line, \citet{marrone2015} demonstrated that applying $\delta$-SPH 
%(i.e. WC-SPH with an extra diffusive term)
the simulation converges to the expected energy loss, while however not being an instantaneous process anymore.
In contrast, in this paper an energy conserving impact is sought, in such a way that kinetic and compressibility energies are exchanged without any net loss, due to the lack of dissipative terms.

%% file: 41_numerical_scheme.tex
\subsubsection{Numerical scheme}
\label{sss:application:scheme}
To carry out the simulations, $L = H = 1$ dimensions are chosen, as well as $\rho_0 = 1$ reference density and $U = 1$ initial velocity magnitude.
No background pressure or volumetric forces are included: $p_0 = 0$,  $\bs{g} = 0$.
No extra dissipation is considered, either artificial viscosity or $\delta$-SPH.
The speed of sound is set at $c_0=100$, as in Ref. \citep{marrone2015}, in order $\text{Ma}\approx 0.01$, so that the system is close to incompressible. Results are largely independent of this value as long as it is
large enough.
A $k = 2$ parameter is set for all the simulations.
Finally, the free-surface is modeled by straight compact support truncation \cite{Colagrossi2009}.

The energy balance in Eq. \eqref{eq:energy:final_balance} can be therefore simplified,
\begin{equation}
\label{eq:practical_application:final_balance}
\dsty{
	P\kin^{n^{\ast}} + P_\mathrm{c}^{n^{\ast}} = P_{\Delta t}(t^{\ast}),
	}
\end{equation}
for some $t^{\ast} \in (t_n, t_{n+1})$.

The term $\SPH{P_{\Grad{\gamma}}}$ has been dropped from the energy balance due to the $k = 2$ parameter selection.
However, $\SPH{P_{\Grad{\gamma}}}$ is not vanishing by any means:
indeed, that term is still able to trade kinetic and internal energy for the sake of exact total energy conservation, as discussed in Sec. \ref{ss:power}.

The terms
$\SPH{P_{\Grad{\gamma}}}$, Eq. \eqref{eq:power:gradgamma}, 
and $P_{\Delta t}(t^{\ast})$ , Eq. \eqref{eq:energy:p_deltat},
do not have a clear sign.
Thus, the corresponding residual are defined with absolute values:
\begin{align}
\label{eq:practical_application:residual_dt}
	R_{\Delta t}(t^{\ast}) = & 
	 \sum_{i \in \Omega} \left\vert
		m_i \overline{\bs{u}_i}(t^{\ast}) \cdot \left(
		\vrate{\bs{u}}{i}{n} - \SPH{\frac{\D \bs{u}}{\D t}}_i(t^{\ast})
	\right) \right\vert  +  \\
	& \nonumber
	\sum_{i \in \Omega} \left\vert
		\frac{m_i \overline{p_i}(t^{\ast})}{\overline{\rho_i}^2(t^{\ast})} \left(
		\vrate{\rho}{i}{n} - \SPH{\frac{\D \rho}{\D t}}_i(t^{\ast})
		\right) \right\vert ,  \\
\label{eq:practical_application:residual_gradgamma}
	R_{\Grad{\gamma}}(t^{\ast}) = & \sum_{i \in \Omega} \left\vert \frac{m_i \overline{p_i}(t^{\ast})}{\overline{\rho_i}(t^{\ast})} \overline{\bs{u}_i}(t^{\ast}) 
	\cdot \SPH{\Grad{\gamma}}_i(t^{\ast}) \right\vert ,
\end{align}
such that the total non-physical energy exchange may be quantified.

In a similar fashion, the following specific energy is defined as well,
%
%\colblue{
\begin{equation}
\label{eq:practical_application:specific_e_gradgamma}
\SPH{e_{\Grad{\gamma}}}_i(t_n) = \Delta t \,\, k\, \sum_{m = 1}^n \left\vert \frac{m_i p_i^m}{\rho_i^m} \bs{u}_i^m \cdot \SPH{\Grad{\gamma}}_i(t_m) \right\vert.
\end{equation}
%}

The fluid domain is discretized in a lattice of $N$ particles with the same mass $m_i = 4 L H / N$, resulting in an initial inter-particle spacing of $\Delta x = 2 L / \sqrt{N}$.
A quintic Wendland kernel \citep{dehnen_aly_wendland_2012} is applied, and its smoothing length $h$ is chosen so that $h / \Delta x = 2$.

%% file: 42_results.tex
\subsubsection{Results}
\label{sss:application:results}
\begin{figure}
	\centering
	\includegraphics[width=0.7\allwidth]{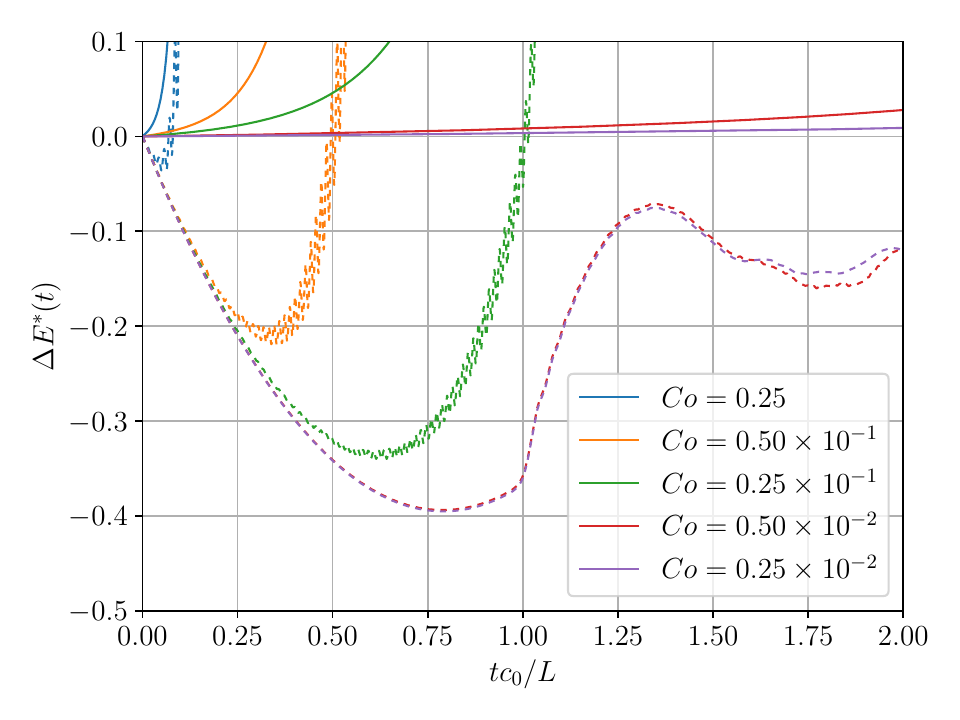}
	\caption{Evolution of relative energy change for different Courant numbers with the explicit Euler scheme. Solid lines: Total energy. Dashed lines: Kinetic energy.}
	\label{fig:application:results:euler:energy}
\end{figure}

Pipelines with the numerical schemes described above have been implemented in the numerical package AQUAgpusph \citep{CercosPita2015, 13th_SPHERIC_Cercos_Calderon_Duque, AQUAgpusph_site}.
In Fig. \ref{fig:application:results:euler:energy} the energy evolution of the system is depicted for the explicit Euler scheme at different Courant numbers.
This is defined as usual, Co=$c_0 (\Delta t) / (\Delta x)$, where
$\Delta x$ is the spatial spacing, which unless otherwise indicated is $L/800$, with the total number of particles set to \added{$N=6.4\times 10^{5}$.}
The variation in energy is quantified by its relative change about its initial value
\begin{equation}\label{eq:energy_metric}
	\Delta E^*(t)  = \frac{E(t) - E(t=0) }{ E\kin(t=0) } ,
\end{equation}
an expression used both for the total energy and the kinetic energy.

\begin{figure}%[!]
	\centering
	\begin{subfigure}{0.45\allwidth}
		\centering
		\includegraphics[width=\textwidth]{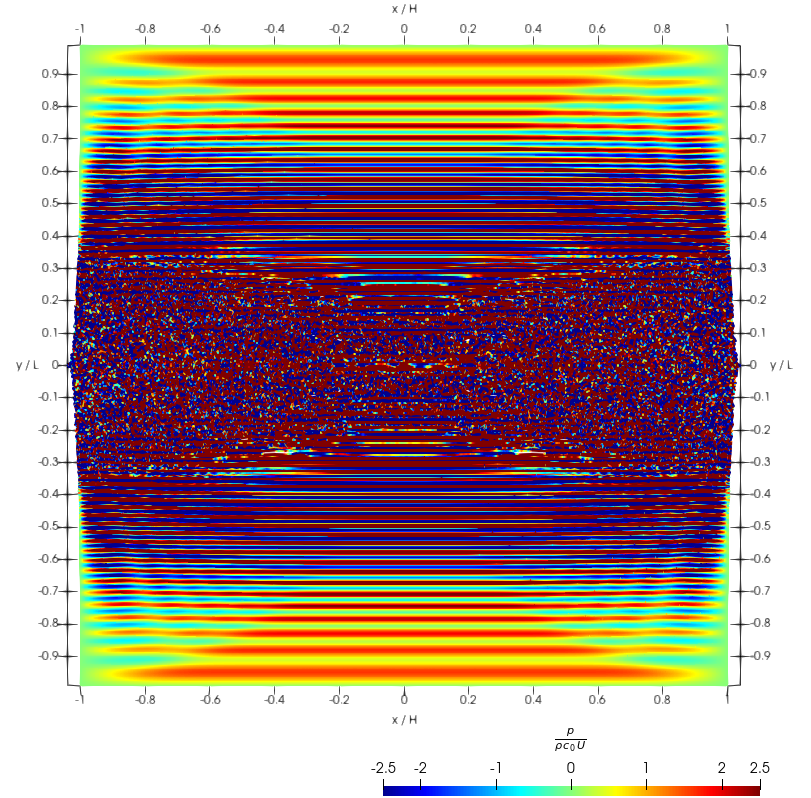}
		\caption{Co $= 0.5 \times 10^{-1}$}
		\label{fig:application:results:euler:snapshots_t1:co=0.1000}
	\end{subfigure}
	~
	\begin{subfigure}{0.45\allwidth}
		\centering
		\includegraphics[width=\textwidth]{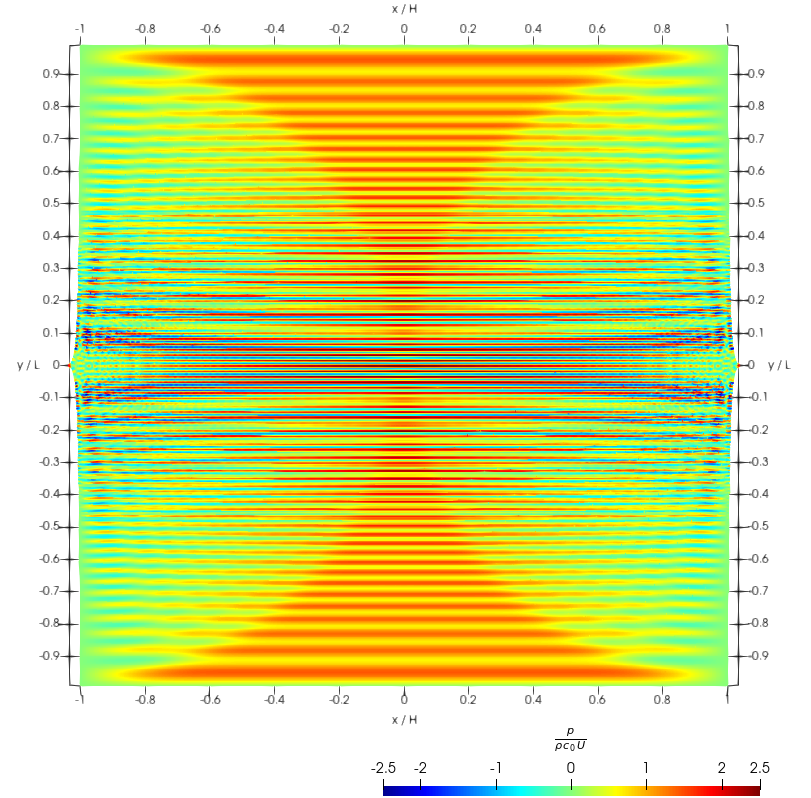}
		\caption{Co $= 0.25 \times 10^{-1}$}
		\label{fig:application:results:euler:snapshots_t1:co=0.0500}
	\end{subfigure}
	\\
	\begin{subfigure}{0.45\allwidth}
		\centering
		\includegraphics[width=\textwidth]{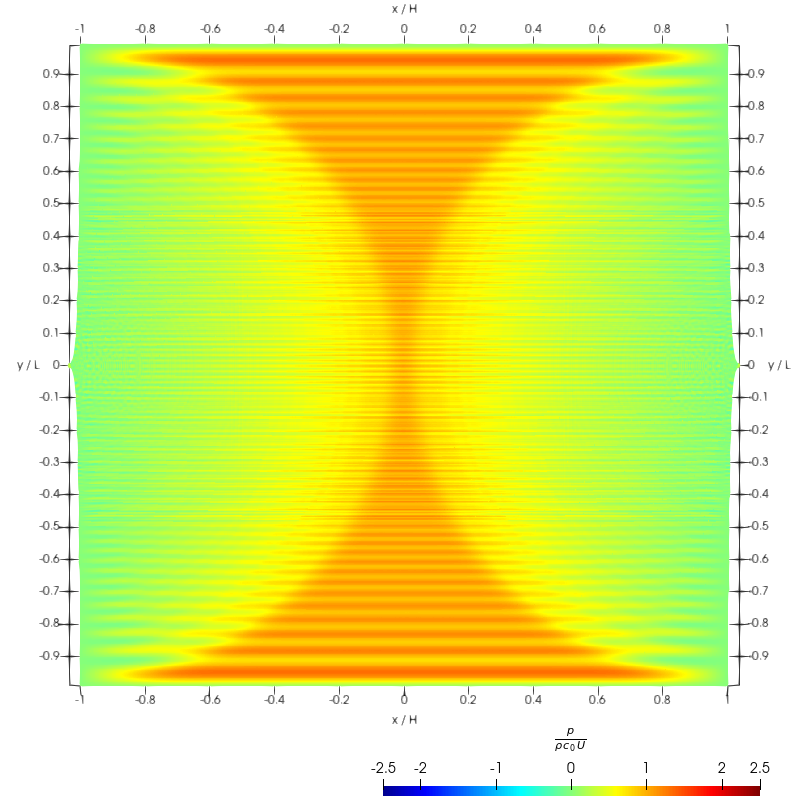}
		\caption{Co $= 0.5 \times 10^{-2}$}
		\label{fig:application:results:euler:snapshots_t1:co=0.0100}
	\end{subfigure}
	~
	\begin{subfigure}{0.45\allwidth}
		\centering
		\includegraphics[width=\textwidth]{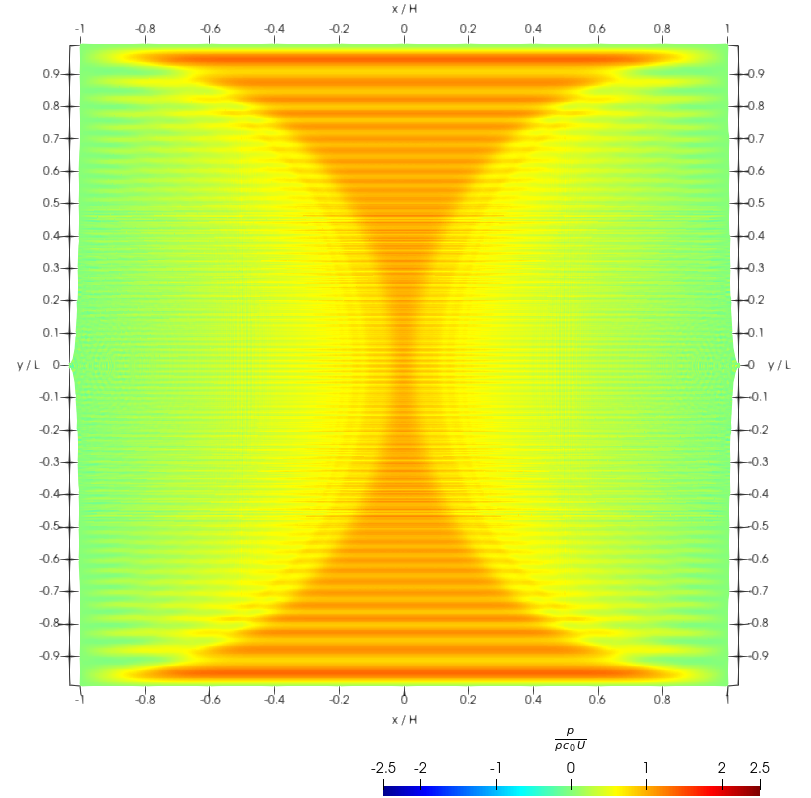}
		\caption{Co$ = 0.25 \times 10^{-2}$}
		\label{fig:application:results:euler:snapshots_t1:co=0.0050}
	\end{subfigure}
	\caption{Snapshots of the non-dimensional pressure field $p/(\rho_0 c_0 U)$ at impact time, $t c_0 / L = 1$, with the explicit Euler scheme.}
	\label{fig:application:results:euler:snapshots_t1}
\end{figure}

As can be appreciated, the simulation quickly turns unstable for large Courant numbers, $\text{Co} \ge 0.25 \times 10^{-1}$.
This is also seen in the simulation snapshots shown in Fig. \ref{fig:application:results:euler:snapshots_t1},
for the ``impact time'' $t = L / c_0 $. This particular time is important, since at this time the pressure shock-wave reaches the top and bottom boundaries. This shock-wave is then reflected as a second shock-wave traveling into the fluid domain, imploding at a time around $t = 2 L / c_0 $, a time which will also be featured in our snapshots, and will be called ``the second impact.''

\begin{figure}
	\centering
	\begin{subfigure}{0.45\allwidth}
		\centering
		\includegraphics[width=\textwidth]{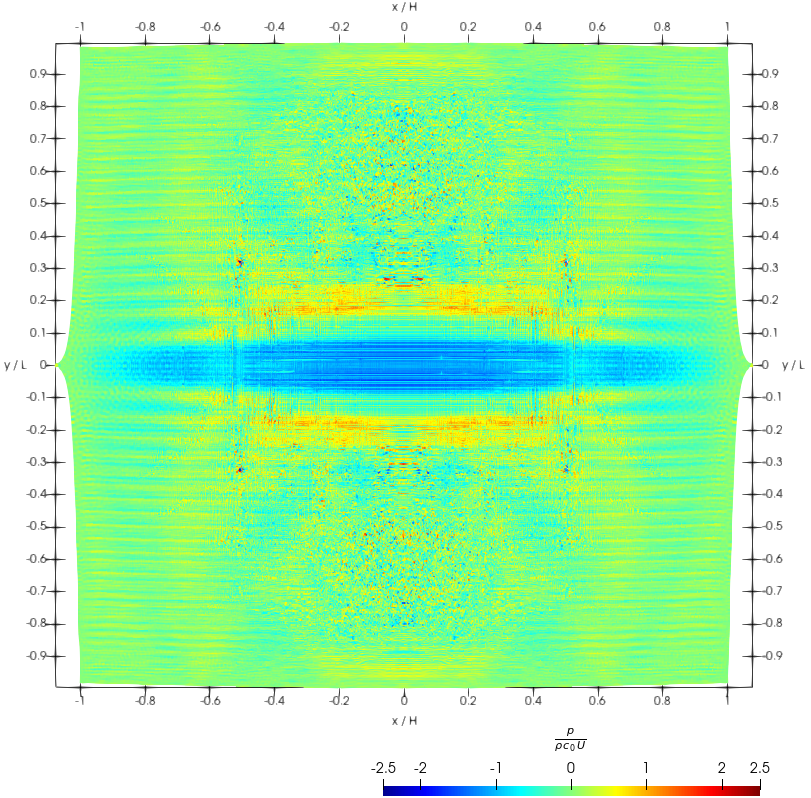}
		\caption{Co = $0.5 \times 10^{-2}$}
		\label{fig:application:results:euler:snapshots_t2:co=0.0100}
	\end{subfigure}
	~
	\begin{subfigure}{0.45\allwidth}
		\centering
		\includegraphics[width=\textwidth]{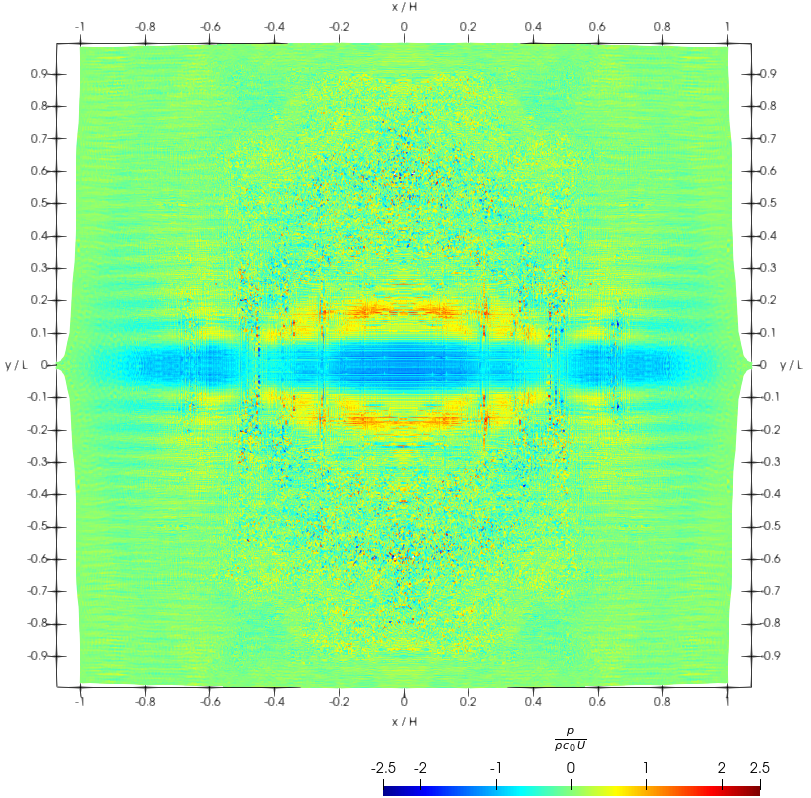}
		\caption{Co = $ 0.25 \times 10^{-2}$}
		\label{fig:application:results:euler:snapshots_t2:co=0.0050}
	\end{subfigure}
	\caption{Snapshots of the simulations at second impact time, $t c_0 / L = 2$, with an explicit Euler scheme.}
	\label{fig:application:results:euler:snapshots_t2}
\end{figure}

For smaller Courant numbers %, $\text{Co} \le 0.5 \times 10^{-2}$,
the total energy is much better conserved.
Some energy is seen in Fig. \ref{fig:application:results:euler:energy} to be artificially pumped into the system anyway, and therefore, the stability of the simulation cannot be granted in the long run.
Moreover, stability is compromised by early spatial instabilities, which resemble the well-known tensile instability. This is indicated by the kinetic energy behavior around the second impact (reduced time around $2$) in same Fig. \ref{fig:application:results:euler:energy}, as well as in the snapshots shown in Fig. \ref{fig:application:results:euler:snapshots_t2} for this time.

\begin{figure}
	\centering
	\includegraphics[width=0.7\allwidth]{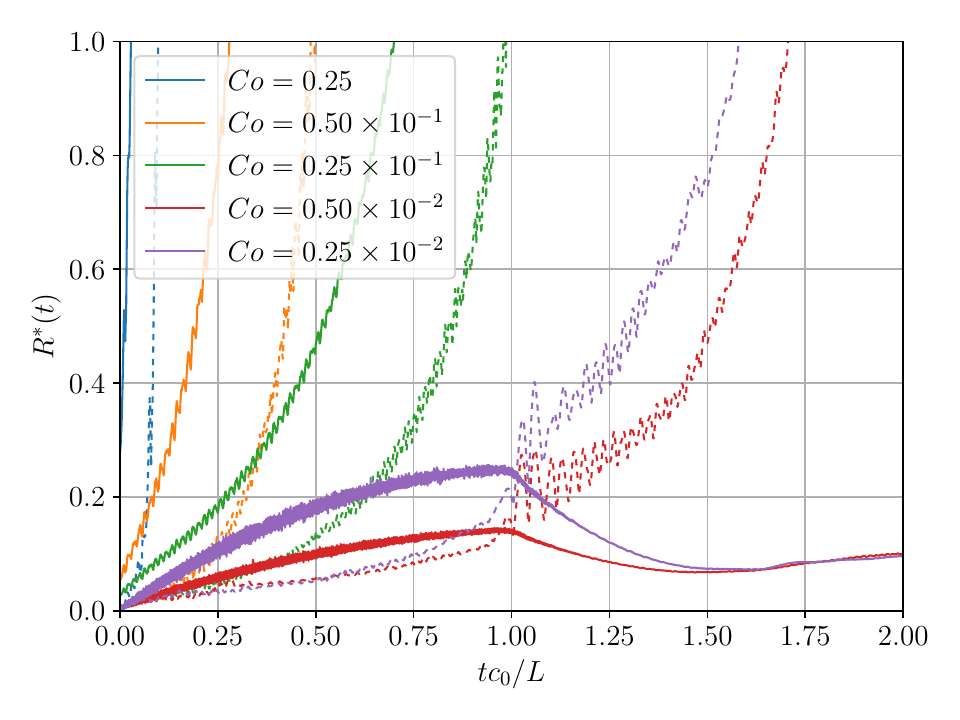}
	\caption{Evolution of residues along the simulation for different Courant numbers with an explicit Euler scheme. Solid lines: $R_{\Delta t}$. Dashed lines: $R_{\Grad{\gamma}}$.}
	\label{fig:application:results:euler:residues}
\end{figure}

The effect of these instabilities is also highlighted in Fig. \ref{fig:application:results:euler:residues}, where the residuals, $R_{\Delta t}$, Eq. \eqref{eq:practical_application:residual_dt},  and $R_{\Grad{\gamma}}$, , Eq. \eqref{eq:practical_application:residual_gradgamma}, are depicted along the simulation. For convenience, these are plotted in
non-dimensional form,
\begin{equation}\label{eq:reduced_residuals}
	R_{\Delta t }^* (t) =  \frac{ R_{\Delta t }(t)  L }{  c_0 E\kin(t=0) } , \qquad
	R_{\Delta \gamma }^*(t) =  \frac{ R_{\Delta \gamma }(t) L }{  c_0 E\kin(t=0) } , \qquad	 
\end{equation}
since the residuals have dimensions of power.
Indeed, for the three largest time steps the term $R_{\Delta t}$ quickly diverges, driving the instability.
Even when the time step is low enough to avoid exponential total energy grow, the phenomenon seems to be dominated by $R_{\Delta t}$ at the early stages of the simulation.
This trend changes after the shockwave reaches the top and bottom boundaries, at the impact time $t c_0 / L = 1$, when the $R_{\Grad{\gamma}}$ residual term becomes the leading one.
As the shockwave is reflected back into the fluid domain, negative pressure values appear, which have been widely associated with tensile instability in the past.

\begin{figure}
	\centering
	\includegraphics[width=0.7\allwidth]{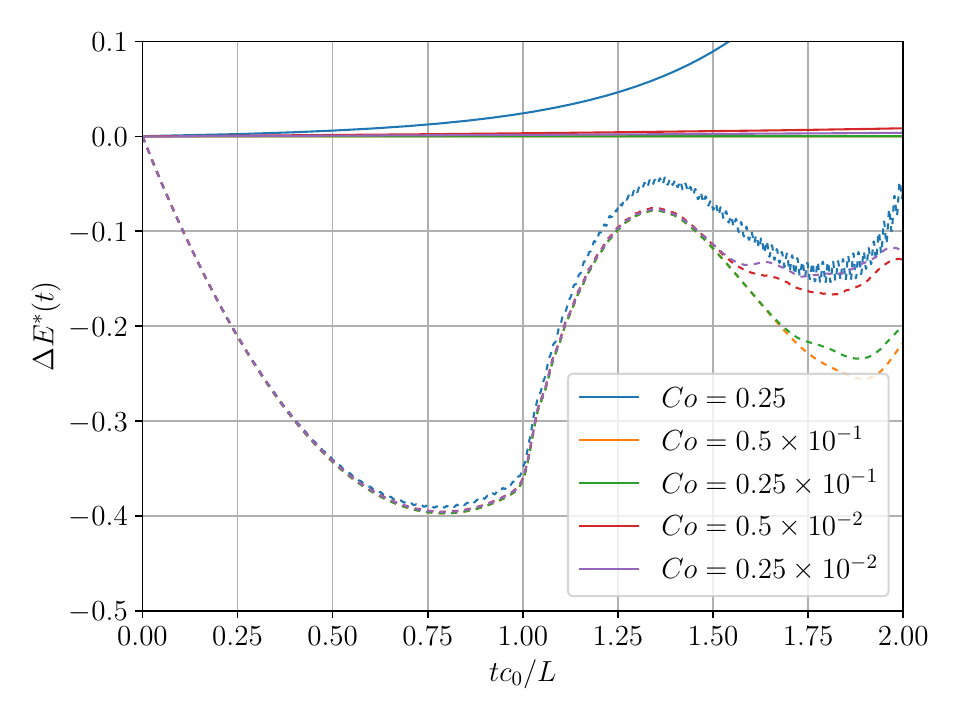}
	\caption{Evolution of relative energy change for different Courant numbers with the explicit Heun method. Solid lines: Total energy. Dashed lines: Kinetic energy.}
	\label{fig:application:results:heuns:energy}
\end{figure}

If the explicit Heun method is employed for these simulations, energy conservation is significantly improved, as can be seen in Fig. \ref{fig:application:results:heuns:energy}.
Only the largest time step causes a significant amount of energy artificially pumped into the system.

\begin{figure}
	\centering
	\includegraphics[width=0.7\allwidth]{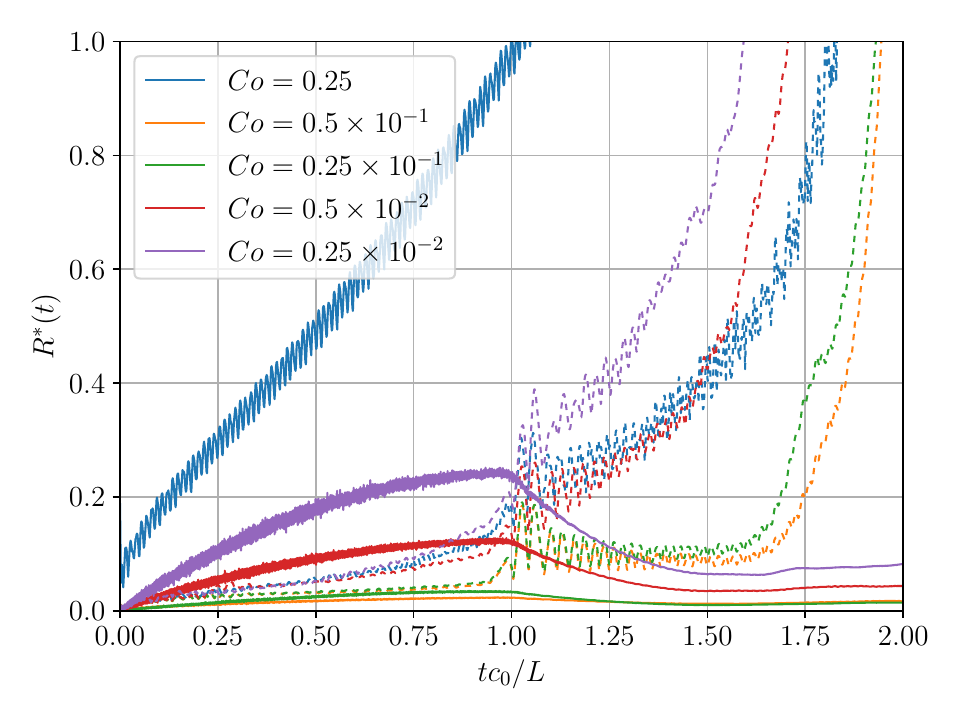}
	\caption{Evolution of residuals along the simulation for different Courant numbers with the explicit Heun method. Solid lines: $R_{\Delta t}$. Dashed lines: $R_{\Grad{\gamma}}$.}
	\label{fig:application:results:heuns:residues}
\end{figure}

Nevertheless, the simulation features trends similar to the explicit Euler method, as shown by the residuals plotted in Fig. \ref{fig:application:results:heuns:residues}.
The simulations again are dominated by tensile instabilities after the shockwave reaches the top and bottom boundaries.

\begin{figure}
	\centering
	\begin{subfigure}{0.45\allwidth}
		\centering
		\includegraphics[width=\textwidth]{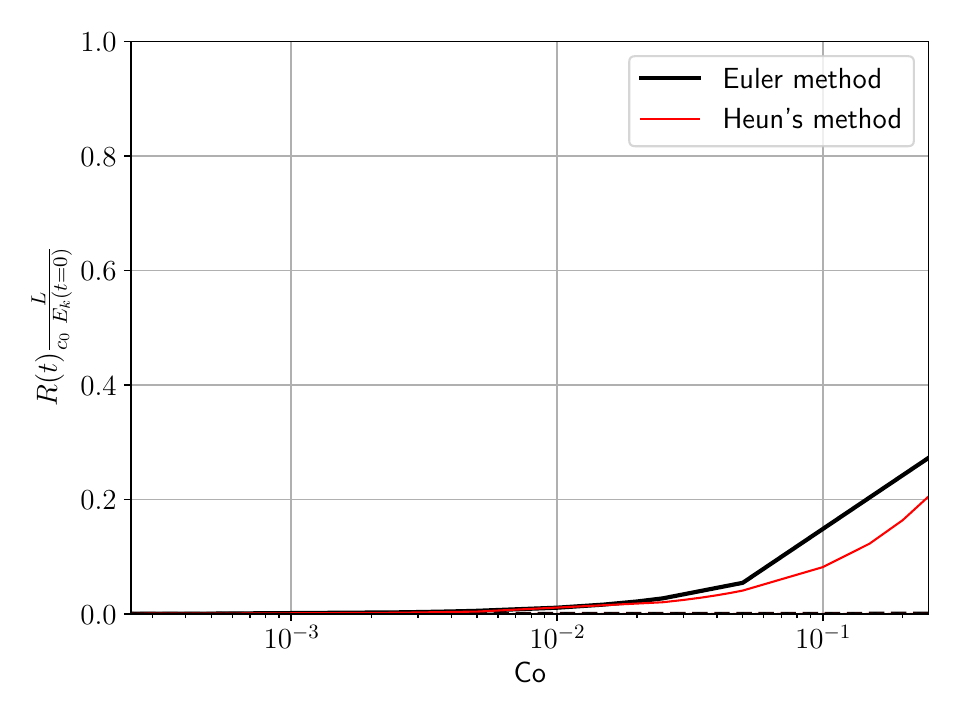}
		\caption{$\frac{t c_0}{L} = 0$}
		\label{fig:application:results:explicit_residues:t0}
	\end{subfigure}
	~
	\begin{subfigure}{0.45\allwidth}
		\centering
		\includegraphics[width=\textwidth]{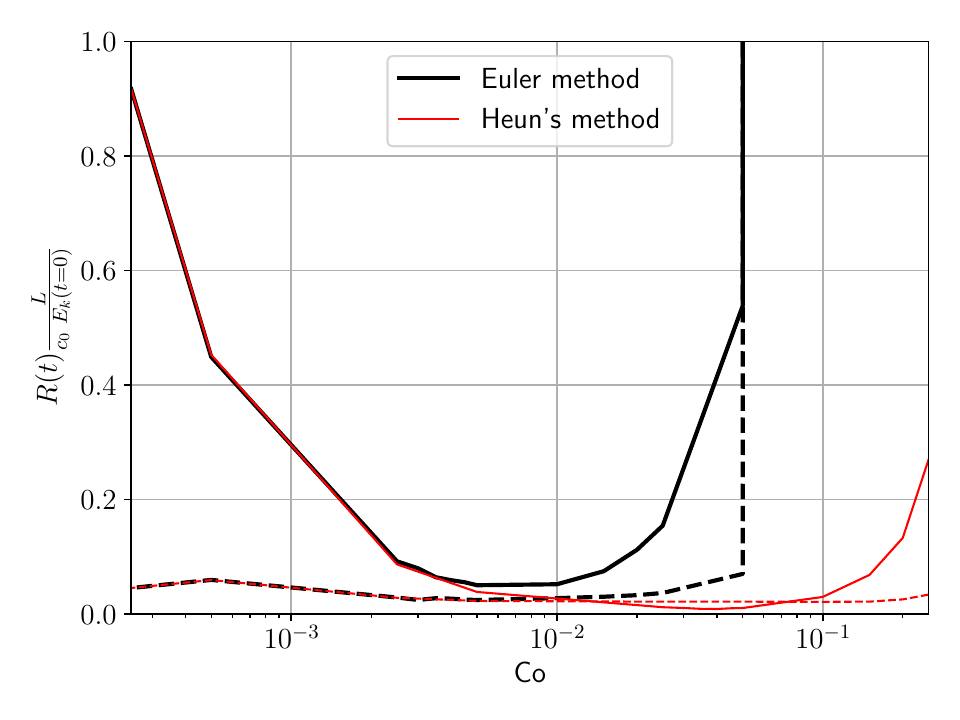}
		\caption{$\frac{t c_0}{L} = 0.2$}
		\label{fig:application:results:explicit_residues:t0.2}
	\end{subfigure}
	\caption{Residues as functions of the Courant number for explicit time integration schemes at 2 different simulation times. Solid lines: $R_{\Delta t}$. Dashed lines: $R_{\Grad{\gamma}}$.}
	\label{fig:application:results:explicit_residues}
\end{figure}

Both explicit time schemes have exposed consistency issues as the time step is decreased.
In fact, in Figs. \ref{fig:application:results:euler:residues} and \ref{fig:application:results:heuns:residues} residues do not seem vanish as the time step is made smaller.
This is made clear in Fig. \ref{fig:application:results:explicit_residues}, where the residues at $t c_0 / L = 0$ and $t c_0 / L = 0.2$ are depicted for different Courant numbers.
Indeed, at the first time step both residues vanish as the time step is diminished, which is not however the case after a short time lapse. The minima in the residues clearly shows an optimal value for the time step, for each time integration scheme.
Further reductions of the time step size inexorably lead to worse energy conservation.
Consistently, the difference between both integration schemes is negligible for the smallest Courant numbers.
%
% The source of such inconsistency is not further analyzed in this paper.
%

\begin{figure}
	\centering
	\includegraphics[width=0.7\allwidth]{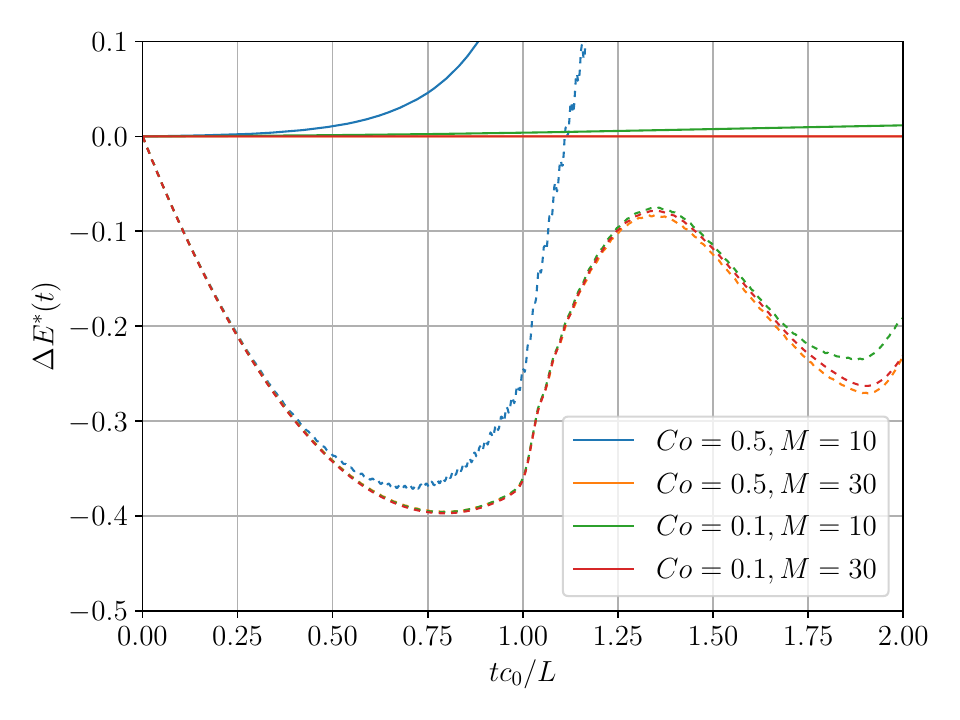}
	\caption{Evolution of relative energy change for different Courant numbers and number of iterations with the implicit midpoint method.
		Solid lines: Total energy. Dashed lines: Kinetic energy.}
	\label{fig:application:results:implicit:energy}
\end{figure}

\begin{figure}
	\centering
	\includegraphics[width=0.7\allwidth]{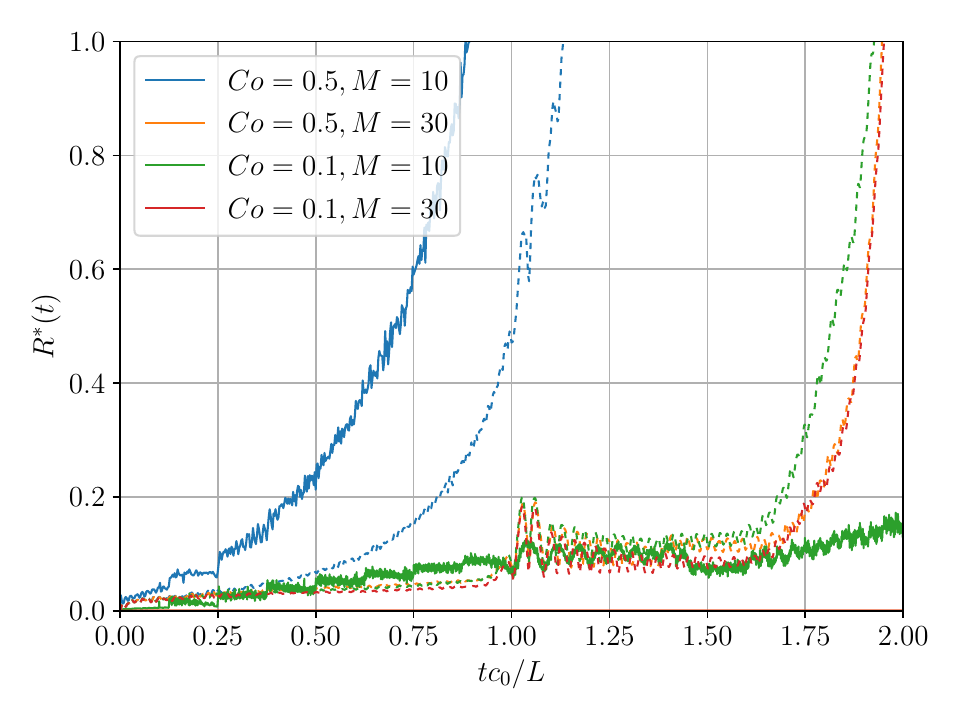}
	\caption{Evolution of residues along the simulation for different Courant numbers and number of iterations when the implicit midpoint method is considered. Solid lines: $R_{\Delta t}$. Dashed lines: $R_{\Grad{\gamma}}$.}
	\label{fig:application:results:implicit:residues}
\end{figure}

Employing the implicit midpoint method, the energy and residues evolution depicted in Figs. \ref{fig:application:results:implicit:energy} and \ref{fig:application:results:implicit:residues} are obtained.

Even for the largest time step, good total energy conservation can be achieved when $30$ iterations are considered to solve the fixed point problems.
Furthermore, it can be appreciated that an increase in the number of iterations computationally outperforms a reduction of the Courant number.

\added{
A more quantitative approach is to describe the number of numerical iterations, $\text{\#s}$, needed in order to evolve the simulation to a given time. The main point is that each iteration has nearly the same computational cost for the explicit and implicit methods, and, while the actual running time will depend on the architecture used, the number of iterations will not. Some estimates of these numbers are now discussed, since we think that an in-deep analysis of computational cost is not so useful for the idealized test cases in this work.}

\added{
In general, $\text{\#s} = t / (\Delta t) $, which in terms of the Courant number can be written as $\text{\#s} = c_0 t / ( \text{Co} \Delta x) $.}

\added{
In most of the simulations of the frontal jet impact of Sec. \ref{ss:normal_jets}, the number of particles is $N=800^2$, and since $L=c_0=1$, we simply have $\text{\#s} = 800 t / \text{Co} \times M $, where $t$ is the physical time, and $M$ the number of inner iterations ($M=1$ for explicit methods). This way, results for the explicit Euler scheme in Figs. \ref{fig:application:results:euler:energy} and \ref{fig:application:results:euler:residues} show that about $160000$ iterations are needed in order to obtain acceptable results at the impact time ($t=1$.)  This number is greatly reduced, to about $16000$, in the Heun method, as seen in Figs. \ref{fig:application:results:heuns:energy} and \ref{fig:application:results:heuns:residues}. These two values are actually close to the optimum ones at $t c_0 / L = 0.2$ in Fig. \ref{fig:application:results:explicit_residues} (b), namely $\text{\#s} / t \approx 160000 $ for the explicit Euler scheme, and $\text{\#s} / t \approx 160000 $ for Heun method.
}

\begin{figure}
	\centering
	\includegraphics[width=0.7\allwidth]{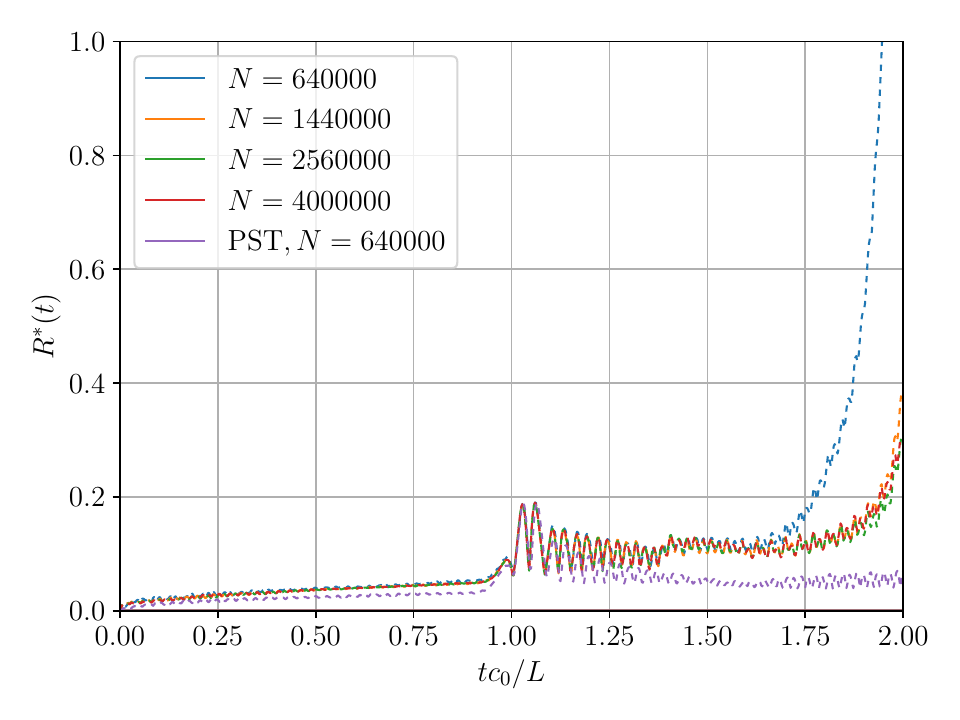}
	\caption{%
Evolution of residues for different number of particles, $N$, at constant Courant number Co $= 0.5$ and number of iterations $M = 30$, with the implicit midpoint method. Solid lines: $R_{\Delta t}$. Dashed lines: $R_{\Grad{\gamma}}$ (numerical results are so close to zero that are not visible at this scale.)
Includes preliminary results from the Particle Shifting Technique to be described in Sec. \ref{s:future_work}.
\label{fig:application:results:implicit:residues_n}
	}
\end{figure}

\added{
The implicit midpoint method, with results in Figs.~\ref{fig:application:results:implicit:energy}, \ref{fig:application:results:implicit:residues}, and \ref{fig:application:results:implicit:residues_n}, seem to require about $48 000$ iterations. According to this metrics, it would then seem that certain explicit methods are actually preferable to our implicit proposal.This is not the case in the next simulation case, the Taylor-Green vortex, as will be discussed in section \ref{ss:taylor_green}.
}

Despite the significantly better energy conservation, the residue associated with a spatial instability, $R_{\Grad{\gamma}}$, is inconsistently converging to a non-zero value when either the time step is decreased or the number of iterations is increased.
A large amount of internal and kinetic energy is artificially exchanged by this term at the second impact, for all considered Courant numbers and number of iterations.
Such non-physical energy exchange can be however diminished by including more particles, as can be seen in Fig. \ref{fig:application:results:implicit:residues_n}, where the residues are plotted for a varying number of particles, $N$, at constant Courant number $\text{Co} = 0.5$, and $M = 30$ iterations.
Negligible total energy residues, $R_{\Delta t}$, are consistently obtained for all the simulations. This permits these simulations to run for long times, without breaking down at the early stages. The figure also features preliminary results from a Particle Shifting Technique which will be briefly described in Sec. \ref{s:future_work}, and show an even better evolution of the residues.

Nevertheless, the $R_{\Grad{\gamma}}$ term is not converging to zero, either in time or in space.
This, is in fact, expected due to the inconsistencies related to the free-surface, as already described by \citet{Colagrossi2009}.

\begin{figure*}[!]
	\centering
	\centering
	\includegraphics[width=0.5\textwidth]{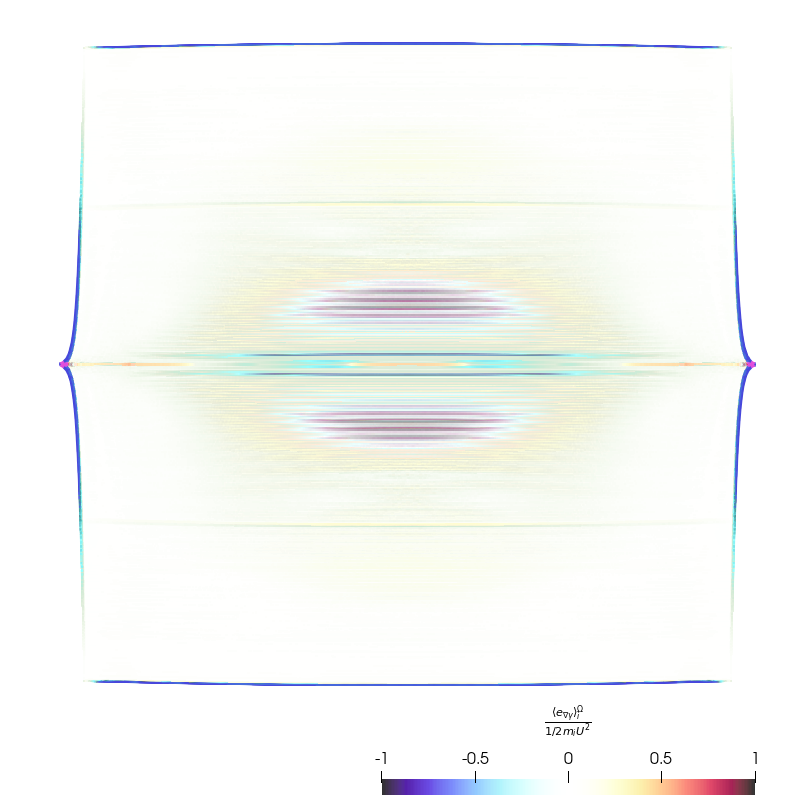}
	\caption{Snapshot of the specific internal energy artificially converted to kinetic energy at time instant $t c_0 / L = 2$. implicit midpoint method, $n = 4\times 10^6$, Co $= 0.5$ and $M = 30$.}
	\label{fig:application:results:implicit:snapshots_t2:co=0.5000}
\end{figure*}

In Fig. \ref{fig:application:results:implicit:snapshots_t2:co=0.5000} the specific internal and kinetic energy artificially traded is depicted for each particle, at the second impact time $t c_0 / L = 2$.
A significant amount of kinetic energy is indeed non-physically transformed into internal energy close to the free-surface.
However, artificially exchanged energy can also be found far away from the free-surface, specially close to the impact line, where largest negative pressure values occur.

%% file: 43_taylor_green.tex
\subsection{Taylor-Green vortex}
\label{ss:taylor_green}
\begin{figure}
	\centering
	\includegraphics[width=0.49\allwidth]{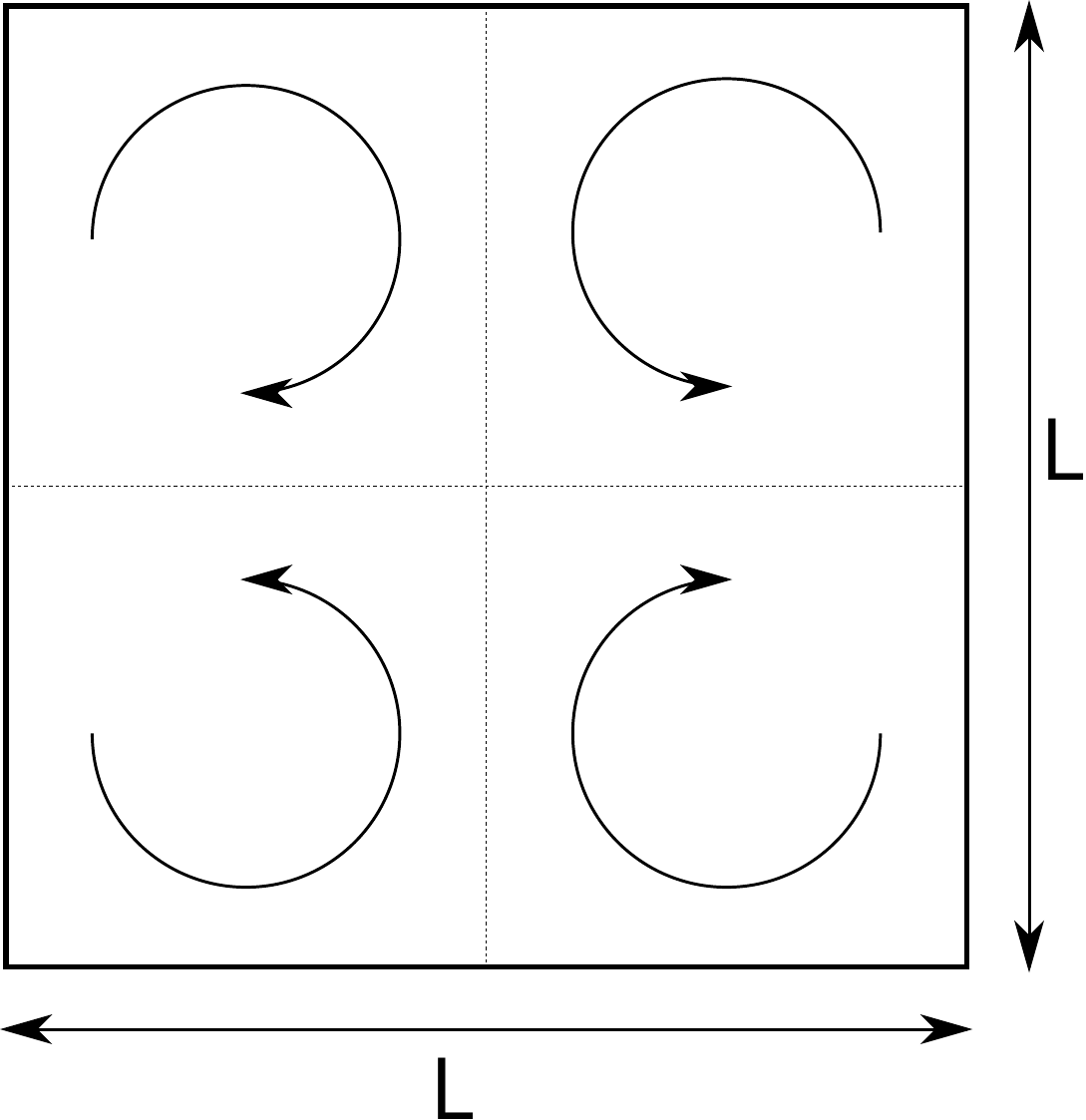}
	\caption{Schematic view of the initial condition for the Taylor-Green vortex simulation.}
	\label{fig:taylor_green:scheme}
\end{figure}

The Taylor-Green Vortex \citep{taylor_1937_taylorgreenvortex} is a popular practical application in which vortices are simulated in a periodic domain. The initial condition is schematically depicted in Fig. \ref{fig:taylor_green:scheme}.
A square domain of size $L \times L$ contains four counter-rotating vortices, described by the initial velocity field
\begin{equation}\label{eq:TG_velocity}
\begin{aligned}
u_x(x, y) = & U_0 \sin \left(\frac{2 \pi}{L} x\right) \cos \left(\frac{2 \pi}{L} y\right)\\
u_y(x, y) =- & U_0 \cos \left(\frac{2 \pi}{L} x\right) \sin \left(\frac{2 \pi}{L} y\right),
\end{aligned}
\end{equation}%
where $U$ is the characteristic velocity.
Periodic boundary conditions are considered.

The corresponding pressure field is
\begin{equation}
	\dsty{
		p(x, y) = p_0 + \frac{\rho_0 U^2}{4} \left( \cos \left(\frac{4 \pi}{L} x\right) + \cos \left(\frac{4 \pi}{L} y\right) \right),
	}
\end{equation}
with $\rho_0$ areplacednd $p_0$ the reference density and background pressure. The time evolution is analytically solvable, with the same expressions as above, but with $U_0$ replaced by $U_0\exp(-2\nu (2\pi /L )^2 t)$, where $\nu$ is the kinematic viscosity coefficient.

In this particular case a Reynolds number, $\text{Re} := L U_0 / \nu = 1000$, is chosen.
As discussed below, larger Reynolds numbers can safely be considered when the implicit time integrator is applied.
However, increasing the Reynolds number for the Euler explicit time scheme would require a extremely large number of particles, as well as an extremely small time step.

%% file: 44_numerical_scheme.tex
\subsubsection{Numerical scheme}
\label{sss:taylor_green:scheme}
To carry out the simulations, $L = 2 \pi$, $\rho_0 = 1$ and $U_0 = 1$ parameters are selected.
No volumetric forces are included, $\bs{g} = 0$. However, in contrast to the previous practical application, a non-null background pressure is considered, $p_0 = 3 \rho_0 U^2$.
The speed of sound is set at $c_0 = 50$, resulting in a $\text{Ma}\approx 0.02$.

The particles are distributed in a lattice grid, with $600$ particles along both the $x$ and $y$ direction. That makes a total of $n=360000$ particles, with initial inter-particle spacing $\Delta x = 2 L / \sqrt{N}$.
At the beginning of the simulation, the particles in the spaces between vortices will compress in one direction and expand in the other. Therefore, an anisotropic distribution of particles will be obtained, which will make the simulation unstable unless the kernel length is large enough.
\added{With the choice of the inter-particle distance over smoothing length ratio made in this particular example, $h = 8 \Delta x$, it follows that the $\alpha$ value representing the artificial viscosity, see \cite{monaghan_arfm_2012, mon2005}, is}
%
%\colblue{
\begin{equation}
	\dsty{
		\alpha = 8 \frac{\text{Ma}}{\text{Re}} \frac{L}{h} = 1.2 \times 10^{-2},
	}
\end{equation}
%}
\added{
a value usually considered low enough to get stable simulations.}

For the sake of simplicity, during this practical simulation only the explicit Euler scheme and the implicit midpoint method are discussed.
Indeed, due to the required small time steps, the differences between the explicit Heun method and the Euler one have been seen to be negligible.

%% file: 45_results.tex
\subsubsection{Results}
\label{sss:taylor_green:results}
\begin{figure}%[!]
	\centering
	\begin{subfigure}{0.45\allwidth}
		\centering
		\includegraphics[width=\textwidth]{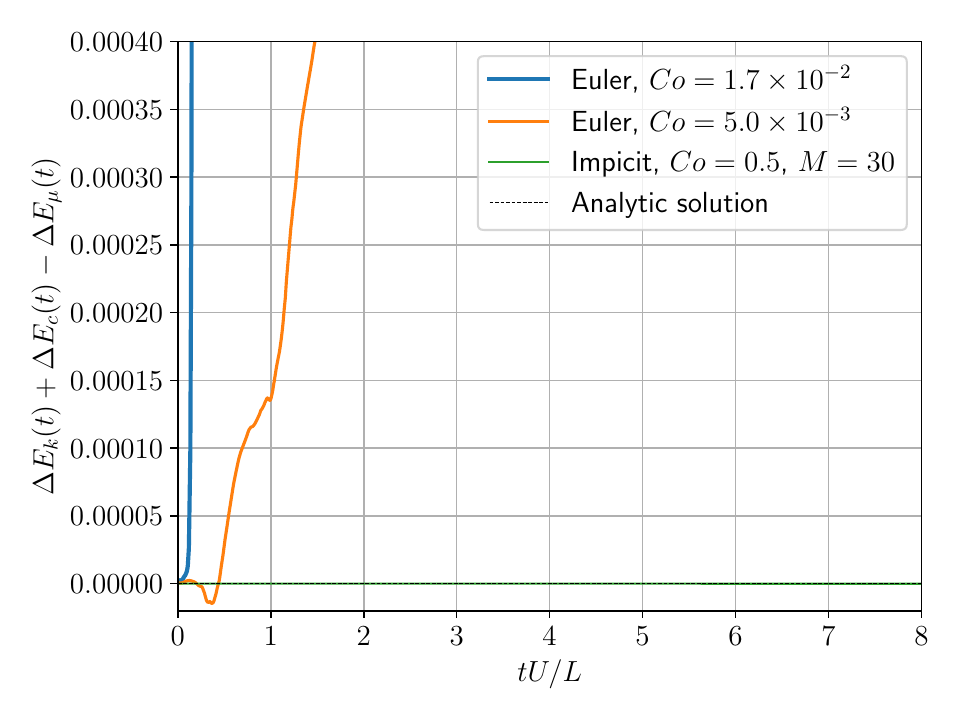}
		\caption{Total energy}
		\label{fig:taylor_green:results:euler:energy_total}
	\end{subfigure}
	~
	\begin{subfigure}{0.45\allwidth}
		\centering
		\includegraphics[width=\textwidth]{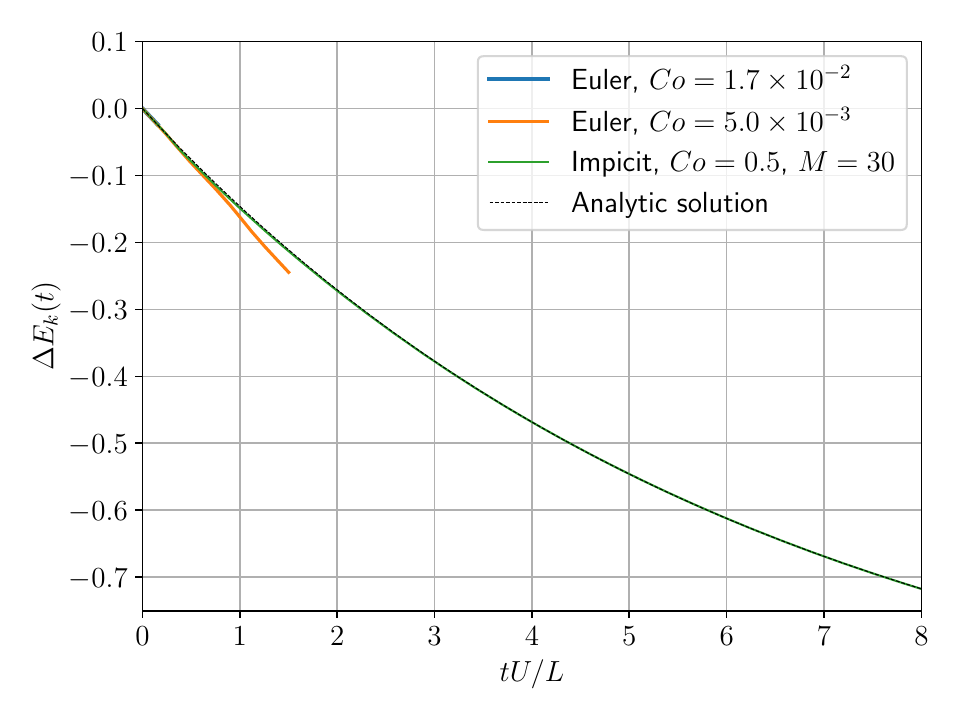}
		\caption{Kinetic energy}
		\label{fig:taylor_green:results:euler:energy_kin}
	\end{subfigure}
	\caption{Evolution of relative energy change for different time integration configurations. The analytical solution is also included.}
	\label{fig:taylor_green:results:euler:energy}
\end{figure}

Simulations are again carried out by the numerical package AQUAgpusph \citep{AQUAgpusph_site, CercosPita2015, 13th_SPHERIC_Cercos_Calderon_Duque}.
In Fig. \ref{fig:taylor_green:results:euler:energy} the energy evolution of the system is depicted for the explicit Euler scheme at different Courant numbers, as well as the implicit time integrator in a unique configuration.
The variation in energy is again quantified by its relative change about its initial value. The total energy includes the kinetic and potential energies, plus the energy dissipated by viscosity (in absolute value), hence its value should remain constant.

As can be appreciated, the implicit midpoint method is able to conserve the total energy of the system for a long time, while capturing the kinetic energy decay phenomena predicted by the analytic solution \citep{taylor_1937_taylorgreenvortex}.
Conversely, when the explicit Euler scheme is considered, a Courant number $\text{Co} = 0.5 / 30 = 1.7 \times 10^{-2}$ would quickly result in an unstable simulation, with poor energy conservation and an overestimated viscous dissipation.
Therefore it can be asserted that the implicit midpoint method is able to outperform the explicit schemes.

\added{
As analyzed for the previous case, if we compare results for numerical iterations at $t U_0 /L = 0.2$ (see Fig. \ref{fig:taylor_green:results:euler:energy}), we find the following: the explicit scheme needs about $1120$ steps for the higher Courant number, and $3800$ for the lower one, while the implicit scheme needs $1140$, \textit{i.e.} with a number of iterations similar to the coarser explicit scheme, the implicit methods provide much better results. For greater times, the explicit results become much less accurate.  For these simulations, $600^2$ particles are used, $U_0=1$, and $L=2\pi$, so $\text{\#s} = 600/(2\pi) t/ \text{Co} \times M $
}

\begin{figure}%[!]
	\centering
	\begin{subfigure}{0.45\allwidth}
		\centering
		\includegraphics[width=\textwidth]{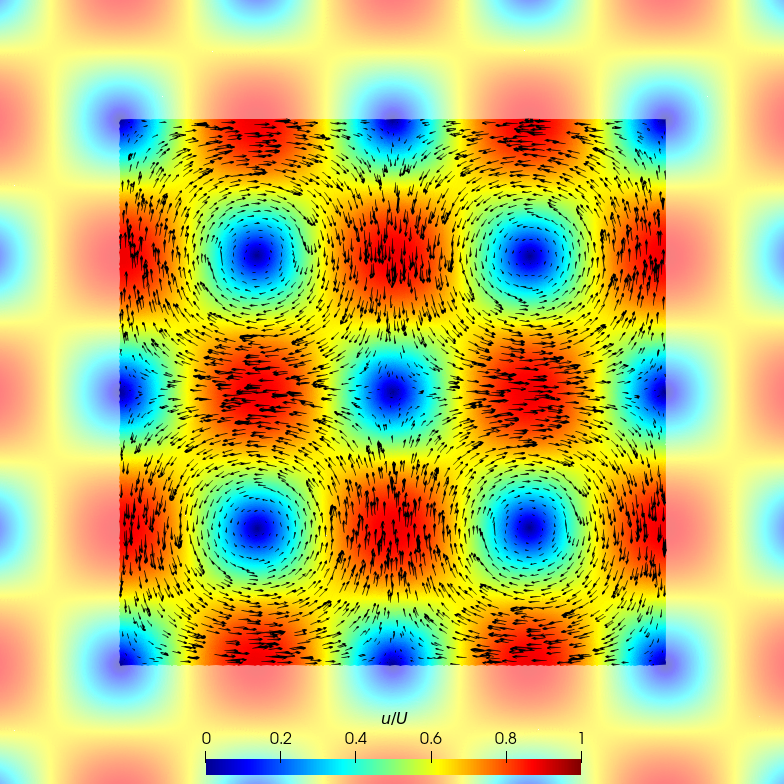}
		\caption{Implicit midpoint time integrator, Co$= 0.5$, $M = 30$}
		\label{fig:taylor_green:results:snapshots_implicit}
	\end{subfigure}
	~
	\begin{subfigure}{0.45\allwidth}
		\centering
		\includegraphics[width=\textwidth]{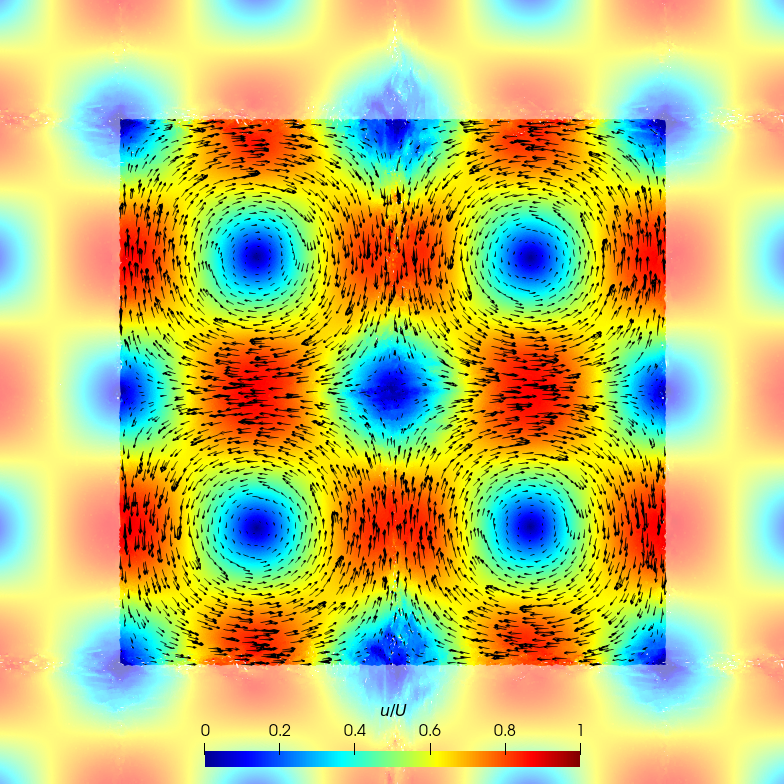}
		\caption{Explicit Euler time integrator, Co$= 5.0 \times 10^-3$}
		\label{fig:taylor_green:results:snapshots_explicit}
	\end{subfigure}
	\caption{Snapshots of the simulations at time instant $t U / L = 1.5$.}
	\label{fig:taylor_green:results:snapshots}
\end{figure}

The simulation is stable for slightly longer if a extremely small Courant number is considered, $\text{Co} = 5 \times 10^{-3}$.
In such case, the explicit Euler scheme is able to keep the energy well conserved for the first half of spin.
Unfortunately it is not able to keep it up, so at the end of the first spin the scheme is already pumping spurious energy in a constant basis.
Those energy conservation flaws induce instabilities, as can be appreciated in Fig. \ref{fig:taylor_green:results:snapshots}.
These noisy fields consequently result in larger kinetic energy dissipation that can be appreciated in Fig. \ref{fig:taylor_green:results:euler:energy}.

Thus, it can be asserted that the implicit midpoint method does not play a role in the viscous term itself, either implying extra numerical dissipation or the inverse.
However, the time integration scheme is indirectly related to viscous dissipation through the noise on the resulting fields.
Along this line, an explicit scheme artificially pumping energy in the system will inexorably result in larger viscous dissipation.

%% file: 50_conclusions.tex
\section{Conclusions}
\label{s:conclusions}

The instantaneous SPH power balance for fluid dynamics has been revisited.
A new power term, associated to spatial instabilities, has been added to the balance.
It has been demonstrated that such term, which was formerly considered as part of the work flow term, can trade kinetic and internal energy, in a non-physical fashion, for the sake of total instantaneous power conservation.

In addition to the improvements on instantaneous power balance, the role of the time integration scheme on the energy balance has also been investigated.
It is demonstrated that energy conservation cannot be granted with explicit time integration schemes, while an implicit midpoint method produces good results (it is likely that other implicit methods are successful in this regard.)
A framework to track the errors on the energy conservation has been discussed,
which may be applied to assessing the quality of simulation results, and to tuning extra dissipation models, like artificial viscosity or $\delta$-SPH.

Simulations of the frontal impact of non-viscous 2-D jets have been carried out.
Non-negligible total energy conservation errors are obtained for all the simulations when either Euler or Heun explicit time integration schemes are considered.
For these, reducing the time step will not monotonically reduce total energy conservation errors, but a minimum error is obtained, for an optimal time step, which depends on the specific scheme.
On the other hand, when the implicit midpoint method is considered, with $30$ inner iterations, negligible total energy conservation errors are obtained, even for large Courant numbers.

\added{Having a look at the number of numerical iterations needed to evolve a simulation to a certain time, the performance of explicit and implicit methods depends on the simulated case. For the inviscid impact test done in the present work, some explicit methods need a smaller number of iterations than the proposed implicit scheme. However, for a case including viscosity, such as the Taylor-Green vortex tested here, the implicit method shows a better performance.}

\added{
Regarding stability, all methods considered fail for the frontal jet impact, the explicit ones around the time of the second impact, and the implicit one around the third one. Preliminary results with a PST show long-term stability for the latter method. For the Taylor-Green vortex, the explicit schemes fail rather quickly, whereas the implicit method is stable throughout the simulation.
}

From a point of view purely centered in practical applications, this is (in our opinion) an important step towards stable WC-SPH simulations, in systems with no extra energy dissipation.
In contrast with previous simulations, in this work kinetic and internal energy are continuously traded as the shockwave travels from the impact front to the fluid edges and back. This leads to simulations that can run for a long time.

The pipelines implemented to carry out the simulations in the SPH solver AQUAgpusph can be found at the project's website \cite{AQUAgpusph_site}.

\section{Future work}
\label{s:future_work}

Although the theoretical analysis may be extended in a straightforward way in order to include energy exchange through the boundaries (see \ref{ss:boundaries}), a practical application in which all those terms are vanishing has been selected. This is because
a wide variety of boundary conditions and formulations should have been otherwise considered.
In addition to that, this energy exchange has already been covered in the literature, with the exception of the term $\left\langle P_{\Grad{\gamma}} \right\rangle^{\partial \Omega}$, Eq.~\eqref{eq:power:work_gradgamma}, an artificial energy exchanged with the boundary which deserves further analysis.

A framework to compute deviations away from total energy conservation has been presented, and applied in our simulations.
However, such framework can be potentially considered to tune extra dissipation terms, like artificial viscosity or $\delta$-SPH, while still applying explicit time integration schemes.
The performance of such approach may be further analyzed in future works.

The simulations show that energy conservation is not converging as the Courant number is reduced, for the explicit time integration schemes.
The roots of that counter-intuitive trend has not been addressed in detail, and should therefore be analyzed in future publications.

A clear avenue for improvement would the alleviation of spatial instabilities, obvious e.g. in Fig. \ref{fig:application:results:euler:snapshots_t1}, perhaps through some particle-shifting technique (PST). In this regard, Fig. \ref{fig:application:results:implicit:residues_n} includes (PST) preliminary results that show a great reduction in the residues, that persists for long simulation times. Details of this PST will shortly be published in a separate
article.

%% file: 60_boundaries.tex
\section{Boundaries}
\label{ss:boundaries}

As mentioned in Eq. \eqref{eq:gov_equations:sph_split}, every single SPH related operator can be split in a
bulk term and a boundary term.

As a consequence, Eq. \eqref{eq:power:balance} would read, in general,
\begin{equation}
	\label{eq:power:balance:boundaries}
	\SPH{P_k}(t) = \SPH{P}^{\partial \Omega}(t) + \SPH{P_{\Grad{p}}}^{\Omega}(t) + \SPH{P_{\mu}}^{\Omega}(t) - P_p(t) +
	k
	\left(
	\SPH{P_{\Grad{\gamma}}}^{\partial \Omega}(t) + \SPH{P_{\Grad{\gamma}}}^{\Omega}(t) 
	\right),
\end{equation}
where the bulk terms are now explicitly marked with an $\Omega$ superscript, and two new
boundary terms appear. One corresponds to the discrete version of the mechanical work done by the boundaries upon the fluid:
\begin{equation}
\label{eq:power:work_k}
	\SPH{P}^{\partial \Omega}(t) = \sum_{i \in \Omega} m_i \bs{u}_i(t) \cdot \left(
	-\frac{\SPH{\Grad{p}}^{\partial \Omega}_i(t)}{\rho_i(t)} + \frac{\mu}{\rho_i(t)} \SPH{\Lap{\bs{u}}}^{\partial \Omega}_i(t)
	\right),
\end{equation}
while the other is due to the $\SPH{\nabla\gamma}$ operator:
\begin{equation}
\label{eq:power:work_gradgamma}
	\SPH{P_{\Grad{\gamma}}}^{\partial \Omega}(t) = -\sum_{i \in \Omega} \frac{m_i p_i(t)}{\rho_i(t)} \bs{u}_i(t) \cdot \SPH{\Grad{\gamma}}^{\partial \Omega}_i(t),
\end{equation}

Eq. \eqref{eq:power:grad_p:balance} likewise has an extra term:
\begin{equation}
	\SPH{P_{\Grad{p}}}^{\Omega}(t) =  - \SPH{P_\mathrm{c}}(t) - 2 \SPH{P_{\Grad{\gamma}}}^{\Omega}(t) + \SPH{P_\mathrm{c}}^{\partial \Omega}(t) , % + \SPH{P_\delta}^{\partial \Omega}(t),	
\end{equation}
where
\begin{equation}
	\SPH{P_\mathrm{c}}^{\partial \Omega}(t) = - \sum_{i \in \Omega} \frac{m_i p_i(t)}{\rho_i(t)}
     \SPH{\Div{\bs{u}}}^{\partial \Omega}_i(t) 	
\end{equation}
is the work exerted by the boundaries on the fluid due to compressibility, already discussed in \citet{CercosPita_etal_CMAME_2017_SPH_ENERGY}.

%	\label{eq:power:work_delta} 
%	\SPH{P_{\delta}}^{\partial \Omega}(t) = \sum_{i \in \Omega} \frac{m_i p_i(t)}{\rho_i^2(t)} \SPH{\frac{\D \rho}{\D t}}^{\delta, \partial \Omega}_i(t).
%
%Above, we have two bulk terms: $\SPH{P_c}^{\Omega}$ is the power due to the compressibility, $\SPH{P_\delta}^{\Omega}$ is the power associated to the $\delta$-SPH-like term, \citep{antuono2015energy}. The other two terms are related to boundaries: $\SPH{P_c}^{\partial \Omega}$ , and $\SPH{P_{\delta}}^{\partial \Omega}$ is the energy exchanged through the boundary possibly arising from a $\delta$-SPH model.

Finally, Eq. \eqref{eq:power:final_balance}, now reads
\begin{align}
	\label{eq:power:final_balance:boundaries}
\SPH{P\kin}(t) & + P_p(t) + \SPH{P_\mathrm{c}}(t) - \SPH{P_{\mu}}^{\Omega}(t) - \SPH{P}^{\partial \Omega}(t)
		= \\
&	\left( k - 2 \right) \SPH{P_{\Grad{\gamma}}}^{\Omega}(t)
+ \SPH{P_\mathrm{c}}^{\partial \Omega}(t)
+ k \SPH{P_{\Grad{\gamma}}}^{\partial \Omega}(t).
\end{align}
Again, all the terms at the right hand side this equation are extra terms, whose presence causes a lack of energy conservation. Conservation is enforced in this work by setting $k=2$, and by considering no boundaries.

The extra work term, $\SPH{P_c}^{\partial \Omega}$, has been already analyzed in previous works \citep{cercospita_thesis_2016, CercosPita_etal_CMAME_2017_SPH_ENERGY}.
On the other hand, the extra work term $\SPH{P_{\Grad{\gamma}}}^{\partial \Omega}$ is new, as it has been traditionally considered part of $\SPH{P}^{\partial \Omega}$.
However, the latter term has a purely numerical nature, and should be therefore split, yielding such an extra term.

It is interesting now to recall the work of \citet{mayrhofer_etal_cpc_2013}, where the possibility of getting $\SPH{\Grad{p}}$ and $\SPH{\Div{\bs{u}}}$ skew-adjoint operators was analyzed within the boundary integrals formulation.
A methodology to achieve that was even proposed therein, but never tested.
It would be therefore interesting to revisit that work to further analyze the relation between the terms $\SPH{P_\mathrm{c}}^{\partial \Omega}$ and $\SPH{P_{\Grad{\gamma}}}^{\partial \Omega}$.
This is however not pursued in this paper, as that will exceed its scope and make it too lengthy.

These terms end up being included in the final balance of Eq. \eqref{eq:energy:final_balance} :
\begin{equation} \label{eq:energy:final_balance:boundaries}
\ldots
	- \SPH{P}^{\partial\Omega}(t_{n + 1/2})    =
\ldots +
	\SPH{P_\mathrm{c}}^{\partial\Omega}(t_{n + 1/2}) +
	\SPH{P_{\Grad{\gamma}}}^{\partial\Omega}(t_{n + 1/2}) ,
\end{equation}
obviously, as extra terms that may cause a lack of energy conservation (for the sake of clarity, the dots stand for all previous terms of Eq. \eqref{eq:energy:final_balance}.)

%% file: 62_hamiltonian.tex
\section{Conservation of energy}
\label{s:hamiltonian}

We briefly show that any two gradient and divergence operators satisfying skew-adjointness property (Eq.~\eqref{eq:power:sym_asym_model}) will lead to the mechanical
energy being conserved in the absence of viscosity. In  \citet{Colagrossi2009} an application of this property within a continuum SPH context can be found.

Euler's momentum equation is \eqref{eq:gov_equations:mom_cons}, neglecting $\mu$ and $\bs{g}$ (the latter, for the sake of simplicity). Considering first $k=2$, it reads
\[
\SPH{\frac{\D \bs{u}}{\D t}}_i(t) =
- \frac{\SPH{\Grad{p}}^\text{sym}_i(t)}{\rho_i(t)}
\]
Applying $m_i \bs{u}_i(t) \cdot$, and summing over all particles,
\[
\sum_i  m_i \bs{u}_i(t) \cdot
\SPH{\frac{\D \bs{u}}{\D t}}_i(t) =
- \sum_i
\frac{ m_i }{\rho_i(t)}
\bs{u}_i(t) \cdot \SPH{\Grad{p}}^\text{sym}_i(t).
\]
This may be written as
\[
\SPH{\frac{\D T}{\D t}}(t) 
 =
- \sum_i
\frac{ m_i }{\rho_i(t)}
\bs{u}_i(t) \cdot \SPH{\Grad{p}}^\text{sym}_i(t) ,
\]
where $T=(1/2)\sum_i m_i u_i^2(t)$ is the total kinetic energy. (In this derivation, time derivatives are supposed to satisfy the usual rules of calculus. This is a subtle point on which we do not have space to dwell here.)

Thanks to skew-adjointness,
\[
\SPH{\frac{\D T}{\D t}}(t) 
=
\sum_i
\frac{ m_i }{\rho_i(t)}
  p_i(t)  \SPH{\Div{\bs{u}}}_i(t) .
\]

Now the appearance of the divergence reminds us of continuity, Eq.
\eqref{eq:gov_equations:mass_cons}, from which
\[
\SPH{\Div{\bs{u}}}_i(t) = -\frac{1}{\rho_i(t)}
\SPH{\frac{\D \rho}{\D t}}_i(t) 
= \rho_i(t) \SPH{\frac{\D (1 / \rho ) }{\D t}}_i(t) .
\]

Therefore,
\[
\SPH{\frac{\D T}{\D t}}(t) 
=
\sum_i
 m_i p_i(t) \SPH{\frac{\D (1 / \rho ) }{\D t}}_i(t) =
\SPH{\frac{\D }{\D t}  \sum_i
	\frac{m_i p_i(t)  }{ \rho_i(t) } }
\]

Then, defining the total internal energy
\[
U(t) = -\sum_i \frac{m_i p_i(t)  }{ \rho_i(t) } ,
\]
We find
\[
\SPH{\frac{\D (T + U)}{\D t}}(t) = 0,
\]
expressing the fact that the mechanical energy, $T+V$, does not change.
If value different than $k=2$ is considered, the final expression is not seen to conserve energy in general. For reference, the final expression using
our antisymmetric operators is
\[
\SPH{\frac{\D T}{\D t}}(t) 
= -
\frac12
\sum_{i,j}
\frac{m_i m_j}{\rho_i(t) \rho_j(t)}
\left[ 
\left(
p_j - p_i
\right)\left(
\bs{u}_i + \bs{u}_j\right)
+
k \left(
p_i \bs{u}_i - p_j \bs{u}_j
\right)
\right]
\cdot \gradient W_{ij} .
\]
If $k=2$, the term inside the square brackets simplifies
\[
\left(
p_j - p_i
\right)\left(
\bs{u}_i + \bs{u}_j\right)
+
2 \left(
p_i \bs{u}_i - p_j \bs{u}_j
\right)
=
\left(
p_i + p_j
\right)\left(
\bs{u}_i - \bs{u}_j\right) ,
\]
and the continuity equation may be used in order to deduce energy conservation.

%% file: 61_delta.tex
\section{Consequences of a $\delta$-SPH model}
\label{ss:deltaSPH}

We briefly discuss, for convenience to interested readers, the modifications that a $\delta$-SPH
formulation would bring about.

In this framework, the conservation Eq. \eqref{eq:gov_equations:mass_cons} has an extra term:
\begin{equation}
	\label{eq:gov_equations:mass_cons:delta}
	\SPH{\frac{\D \rho}{\D t}}_i(t) = -\rho_i(t) \SPH{\Div{\bs{u}}}_i(t) +
	\SPH{\frac{\D \rho}{\D t}}^\delta_i(t),
\end{equation}
whose specific shape we leave unspecified, for the sake of generality.

Eq. \eqref{eq:power:balance:boundaries} does not change, but Eq. \eqref{eq:power:final_balance:boundaries} does:
\begin{equation}
\ldots =  \ldots   +  \SPH{P_\delta}^\Omega(t) 	+  \SPH{P_\delta}^{\partial \Omega}(t) 	,
\end{equation}
where the dots represent, again for simplicity, the part of Eq. \eqref{eq:power:final_balance:boundaries} that has not changed, and:
\begin{equation}
\SPH{P_\delta}^\Omega(t) = 
\sum_{i\in\Omega} \frac{m_i p_i(t)}{\rho^2_i(t)}  \SPH{\frac{\D \rho}{\D t}}^{\delta, \Omega}_i (t) ,
\qquad
 \SPH{P_\delta}^{\partial \Omega}(t)  =  
\sum_{i\in\Omega} \frac{m_i p_i(t)}{\rho^2_i(t)}  \SPH{\frac{\D \rho}{\D t}}^{\delta, \partial\Omega}_i (t) .
\end{equation}
I.e. the $\delta$-SPH term gives rise to the first, bulk power, term, \citep{antuono2015energy} and the second term, related to the energy exchanged through the boundaries.

\citet{antuono2015energy} previously merged $\SPH{P_\delta}^{\Omega}$ and  $\SPH{P_{\delta}}^{\partial \Omega}$ into a single term, but we propose it should be split.
This allows to relax the conditions imposed on $\SPH{P}^{\Omega}_\delta$, in such a way that the theoretical analysis in that work becomes valid --- in particular, the fact that the term leads to energy dissipation, therefore contributing to the stability of the model.
They also analyzed numerically the overall contribution from both terms, checking that they were not pumping energy in the system, so it can be expected that the effect of the $\SPH{P_{\delta}}^{\partial \Omega}$ term is negligible.
A formal analysis is still required though.

Equation \eqref{eq:power:final_balance:boundaries} would now include these two terms:
\begin{equation}
	\ldots =
	\ldots +
	\SPH{P_\delta}^{\partial \Omega}(t) +
	\SPH{P_\delta}^{\partial \Omega}(t) ,
\end{equation}
on its right-hand, thus representing a deviation from energy conservation. Accordingly, \eqref{eq:energy:final_balance:boundaries} would now feature these terms:
\begin{equation}
	\ldots
=	\ldots +
	\SPH{P_\delta}^{\partial \Omega}(t_{n+1/2}) +
	\SPH{P_\delta}^{\partial \Omega}(t_{n+1/2}) 
\end{equation}